\documentclass[aps,prd, twocolumn,preprintnumbers,nofootinbib,superscriptaddress]{revtex4-2}
\usepackage{amsmath, amssymb, graphicx, setspace,color,url,xcolor}
\usepackage{cases} 				
\usepackage{CJK}              
\usepackage{tikz}
\usetikzlibrary{decorations.pathreplacing,decorations.markings, positioning}
\usepackage{fix-cm}
\usepackage{mathtools}

\newcommand {\eV}          {{\,\rm eV}}
\newcommand {\pc}          {{\,\rm pc}}
\newcommand {\kpc}         {{\,\rm kpc}}
\newcommand {\Myr}         {{\,\rm Myr}}
\newcommand {\Gyr}         {{\,\rm Gyr}}
\newcommand {\Msun}        {{\,\rm{M}_{\odot}}}
\newcommand {\rhosc}       {\rho_{\rm SC}}
\newcommand {\rhosol}      {\rho_{\rm c}}
\newcommand {\tsol}        {\tau_{\rm soliton}}
\newcommand {\tSC}         {\tau_{\rm SC}}
\newcommand {\tosc}        {\tau_{\rm osc}}
\newcommand {\Rhl}         {R_{\rm hl}}
\newcommand {\Rhli}        {R_{\rm hl,0}}
\newcommand {\MN}          {MN2019}
\newcommand {\Rhlmax}      {R_{\rm hl,max}}
\newcommand {\Wosc}        {\omega_{\rm osc}}

\newcommand {\sref}[1]     {Section~\ref{#1}}
\newcommand {\fref}[1]     {Fig.~\ref{#1}}

\newcommand {\eref}[1]     {Eq.~(\ref{#1})}
\newcommand {\pAeref}[1]     {Eqs.~(\ref{#1})}
\newcommand {\pBeref}[1]     {~and~(\ref{#1})}
\newcommand {\pCAeref}[1]     {Eqs.~(\ref{#1}}
\newcommand {\pCBeref}[1]     {, \ref{#1}}
\newcommand {\pCCeref}[1]     {, \ref{#1})}
\newcommand {\Seref}[1]     {Equation~(\ref{#1})}

\newcommand {\be}          {\begin{equation}}
\newcommand {\ee}          {\end{equation}}

\newcommand{\eee}{\equiv}

\usepackage{scalerel,tikz}
\usetikzlibrary{svg.path}
\definecolor{orcidlogocol}{HTML}{A6CE39}
\tikzset{orcidlogo/.pic={
		\fill[orcidlogocol] svg{M256,128c0,70.7-57.3,128-128,128C57.3,256,0,198.7,0,128C0,57.3,57.3,0,128,0C198.7,0,256,57.3,256,128z};
		\fill[white] svg{M86.3,186.2H70.9V79.1h15.4v48.4V186.2z}
		svg{M108.9,79.1h41.6c39.6,0,57,28.3,57,53.6c0,27.5-21.5,53.6-56.8,53.6h-41.8V79.1z M124.3,172.4h24.5c34.9,0,42.9-26.5,42.9-39.7c0-21.5-13.7-39.7-43.7-39.7h-23.7V172.4z}
		svg{M88.7,56.8c0,5.5-4.5,10.1-10.1,10.1c-5.6,0-10.1-4.6-10.1-10.1c0-5.6,4.5-10.1,10.1-10.1C84.2,46.7,88.7,51.3,88.7,56.8z};
}}
\newcommand\orcidicon[1]{\href{https://orcid.org/#1}{\mbox{\scalerel*{
				\begin{tikzpicture}[yscale=-1,transform shape]
					\pic{orcidlogo};
				\end{tikzpicture}
			}{|}}}}

\begin{document}
\begin{CJK*}{UTF8}{bkai}
\title{Soliton Oscillations and Revised Constraints from Eridanus II of Fuzzy Dark Matter}

\author{Barry T. Chiang~\orcidicon{0000-0002-9370-4490}}
\affiliation{Department of Applied Mathematics and Theoretical Physics, University of Cambridge, Cambridge, CB3 0WA, UK}
\affiliation{\textit{Institute of Astrophysics, National Taiwan University, Taipei 10617, Taiwan}}

\author{Hsi-Yu Schive (薛熙于)~\orcidicon{0000-0002-1249-279X}}
\email{hyschive@phys.ntu.edu.tw}
\affiliation{\textit{Institute of Astrophysics, National Taiwan University, Taipei 10617, Taiwan}}
\affiliation{\textit{Department of Physics, National Taiwan University, Taipei 10617, Taiwan}}
\affiliation{\textit{Center for Theoretical Physics, National Taiwan University, Taipei 10617, Taiwan}}
\affiliation{Physics Division, National Center for Theoretical Sciences, Taipei 10617, Taiwan}
					
\author{Tzihong Chiueh (闕志鴻)~\orcidicon{0000-0003-2654-8763}}
\affiliation{\textit{Institute of Astrophysics, National Taiwan University, Taipei 10617, Taiwan}}
\affiliation{\textit{Department of Physics, National Taiwan University, Taipei 10617, Taiwan}}
\affiliation{\textit{Center for Theoretical Physics, National Taiwan University, Taipei 10617, Taiwan}}
	
	\begin{abstract}
Fuzzy dark matter (FDM) has been a promising alternative to standard cold dark matter.
The model consists of ultralight bosons with mass $m_b \sim 10^{-22}\eV$ and features
a quantum-pressure-supported solitonic core that oscillates. In this work, we
show that the soliton density oscillations persist even after significant tidal stripping of the
outer halo. We report two intrinsic yet distinct timescales
associated, respectively, with the ground-state soliton wavefunction $\tau_{00}$
and the soliton density oscillations $\tau_\text{soliton}$, obeying
$\tau_\text{soliton} /\tau_{00} \simeq 2.3$. The central star cluster (SC)
in Eridanus II has a characteristic timescale
$\tau_\text{soliton} / \tau_\text{SC} \sim 2\textendash3$ that
deviates substantially from unity. As a result, we demonstrate, both analytically
and numerically with three-dimensional self-consistent FDM simulations, that
the gravitational heating of the SC owing to soliton density oscillations is
negligible irrespective of $m_b$. We also show that the subhalo mass function to
form Eridanus II does not place a strong constraint on $m_b$. These results are
contrary to the previous findings by Marsh \& Niemeyer (2019).
	\end{abstract}

	\maketitle
\end{CJK*}
	\section{Introduction}\label{sec:Intro}
	The $\Lambda$CDM paradigm has been demonstrably successful in describing the observed large-scale structure, accounting for the existence of the cosmic microwave background, and naturally explaining the current accelerated expansion of the Universe. Many predictions of the standard cold dark matter (CDM) model on sub-galactic scales, however, appear inconsistent with observations (e.g. the Core-cusp, Missing Satellites, and Too-Big-to-Fail problems) \cite{Bullock:2017xww, Weinberg:2013aya, Spergel:1999mh, Kamionkowski:1999vp, SommerLarsen:1999jx}.

	As a well-motivated alternative DM candidate, the fuzzy dark matter (FDM) \cite{Hu:2000ke, Marsh2016, Hui:2016ltb, Niemeyer:2019aqm} is composed of ultralight bosons with a particle mass $m_b \sim 10^{-22}$\textendash$10^{-20}$ eV. The model offers both observationally consistent descriptions of small-scale structure and large-scale predictions consistent with the $\Lambda$CDM framework \cite{Schive:2014dra, Marsh:2015wka, Calabrese:2016hmp, Chen:2016unw, Gonzales-Morales:2016mkl}. Exhibiting astrophysically large de Broglie wavelength, FDM features a standing wave soliton supported by quantum pressure forming at the center of every galaxy, accompanied by an extensive NFW-like halo \cite{Schive:2014dra, Schive:2014hza}. The rich phenomenology of FDM has been explored actively via numerical simulations \cite{Schwabe:2016rze, Mocz:2017wlg, Lin:2018whl, Veltmaat:2018dfz, Veltmaat:2019hou, Li:2020ryg}. However, despite the substantial progress by simulations in revealing many intriguing aspects of FDM, mechanisms such as soliton coherent density oscillations, stochastic density fluctuations (halo granules), and soliton random walk excursions still require further theoretical investigations and observational support.

	A recent investigation into the survival of the ancient central star cluster (SC) of the ultra-faint dwarf galaxy Eridanus II (Eri II) by Marsh \& Niemeyer (2019) \cite{Marsh:2018zyw}, henceforth abbreviated as \MN, estimated the gravitational heating effect caused by stochastic fluctuations and soliton density oscillations. By approximating the dynamical timescales of both mechanisms, a lower bound $m_b \gtrsim 10^{-19}$ eV (with partial exclusions in the range $10^{-20} < m_b < 10^{-19}$ eV) was derived together with the constraint from the subhalo mass function. This bound is rather stringent compared with other constraints from, for example, Lyman-$\alpha$ forest $m_b \gtrsim 2\times10^{-21}$ eV \cite{Kobayashi:2017jcf, Irsic:2017yje, Armengaud:2017nkf}, DM (sub)halo abundance $m_b \gtrsim 2\textendash3 \times 10^{-21}\eV$ \cite{Nadler:2019zrb, Schutz:2020jox, Nadler:2020prv}, galaxy luminosity function at high redshifts $m_b \gtrsim 1\textendash8\times10^{-22}\eV$ \cite{Schive:2015kza, Corasaniti:2016epp, Menci:2017nsr}, and heating of galactic disks $m_b \gtrsim 1\textendash2\times10^{-22}\eV$ \cite{Church2019, BarOr2019, Amr2020}.

	In this work, we present a theoretical framework that pins down the dynamical timescales of both the ground-state soliton wavefunction $\tau_{00}$ and the coherent density oscillations $\tau_\text{soliton}$ for a perturbed FDM soliton. Concerning the stability of Eri II SC, \cite{Schive:2019rrw} demonstrates that the SC undergoes a complete tidal disruption within $\sim 1\Gyr$ due to soliton random walk excursions in the presence of a background halo; the effect can be largely mitigated if the background halo is stripped away by the Milky Way tides. Observational constraints on Eri II \cite{Crnojevi:2016ApJ, Li:2016utv} indicate that the mass contribution within its half-light radius is predominantly sourced by the soliton core for $m_b \lesssim 10^{-21}$ eV. We confirm both analytically and with N-body and FDM simulations that the effect of gravitational heating on Eri II SC due to soliton density oscillations is negligible for $m_b \lesssim 4\times10^{-21}\eV$, contrary to the conclusion drawn by \MN. We further show that the FDM constraints derived by \MN\, from the subhalo mass function is similarly invalid.

	We begin with an overview on the analytical descriptions of FDM and the astrophysical constraints on the soliton profile from Eri II in Sec. \ref{OscillationFrequencyRevisit}. In Secs. \ref{Heating} and \ref{sec:sim}, we present both analytical and numerical calculations on the SC heating and demonstrate that orbital resonances are inefficient for $m_b \lesssim 4\times10^{-21}$ eV, which is followed by remarks concerning \MN\, in Sec. \ref{app:TragetPaper}. In Sec. \ref{sec:conclude} we conclude. Technical details are organized as follows: derivations of and discussions on relevant features of an unperturbed soliton (Appendix \ref{app:SolitonUnperturbed}), linear soliton density perturbations (Appendix \ref{app:PerturbationAppendix}), gravitational heating of the SC (Appendix \ref{app:HamiltonianPerturbation}), and resonance analysis (Appendix \ref{app:Resonance}).

\section{Fuzzy Dark Matter Solitons: Properties and Constraints}\label{OscillationFrequencyRevisit}
In this section, we examine aspects of the FDM soliton and the astrophysical constraints from Eri II on the soliton density profile. We identify two distinct timescales associated with the ground-state soliton wavefunction and solitary coherent density fluctuations respectively.
\subsection{Soliton Oscillations}\label{sec:SolitonTimeScale}
The evolution of FDM is described by a classical complex scalar $\Psi$ that obeys the coupled Schr\"{o}dinger-Poisson (SP) equations \cite{Hu:2000ke, Hui:2016ltb, Niemeyer:2019aqm}
\begin{eqnarray}
i \hbar \frac{\partial \Psi}{\partial t} &=& -\frac{\hbar^2}{2m_b} \nabla^2 \Psi + m_bV\Psi,\label{eqn:SPeqnS}\\
\nabla^2 V &=& 4\pi G m_b |\Psi|^2, \label{eqn:SPeqnP}
\end{eqnarray}
where $V$ denotes the gravitational potential, $G$ the gravitational constant, and $m_b|\Psi|^2 = \rho_\text{FDM}$ the mass density. The density profile of an unperturbed (ground-state) soliton $\rho_\text{soliton}$ and the characteristic radius of a soliton $r_c$ at which $\rho_\text{soliton}=\rho_c/2$ read \cite{Schive:2014dra}
\begin{eqnarray}
&&\rho_\text{soliton}(\gamma; \rho_c, m_b) =\rho_c\big(1+9.1\times 10^{-2}\gamma^2\big)^{-8},\;\;\;\;\;\;\;\;\;\;\;\;\;\;\;\:\:\:\:\label{eqn:gammaDensityDist}\\
&&r_c(\rho_c, m_b)=\Big(\frac{\rho_c}{1.9 \text{ M}_\odot \text{pc}^{-3}}\Big)^{-1/4} \Big(\frac{m_b}{10^{-23}\text{eV}}\Big)^{-1/2} \text{kpc}, \;\;\;\;\label{eqn:SolitonRadiusEriII}
\end{eqnarray}
where $\rho_c$ denotes the soliton central density and we have introduced a dimensionless spatial variable
\begin{eqnarray}
\gamma &\eee& \frac{r}{r_c}. \label{eqn:ScaledRadius}
\end{eqnarray}

The gravitational potential of a soliton takes the form
\begin{eqnarray}\label{eqn:VExcitedStatesAnalysis}
&&V_0(\gamma; \rho_c, m_b) = -G r_c^2 \rho_c \Bigg\{\bigg[\frac{4.17\times 10^{-17}}{(1+9.1\times 10^{-2}\gamma^2)^6}\bigg]\big(1.83 \\
&&\times 10^{17}+5.14\times 10^{16}\gamma^2 +7.09\times 10^{15}\gamma^4+ 5.30\times 10^{14}\gamma^6 \nonumber \\
&&+2.07\times 10^{12}\gamma^{8}+3.33\times 10^{11} \gamma^{10}\big) + \frac{7.38\arctan{\big(0.302\gamma\big)}}{\gamma}\Bigg\},\nonumber
\end{eqnarray}
which suggests a cutoff radius $\gamma_\text{cut} = 3.3$ (Fig. \ref{fig:SolitonPotential}), beyond which the soliton acts gravitationally as a point mass. Relatedly, $\gamma_\text{cut}$ also offers a natural explanation to the observed universal soliton-halo profile transition radius $\gamma \simeq 3.5$ \cite{Mocz:2017wlg}. The eigenfunctions of an unperturbed soliton $\psi_{nlm}(\gamma, \theta, \phi) = (r_c \gamma)^{-1}u_{nl}(\gamma) \times Y_l^m(\theta, \phi)$ (\eref{eqn:WaveFDecomposition}) are derived in Appendix \ref{app:SolitonUnperturbed}. The associated energy levels $E_{nl}$ scales as $\rho_c^{1/2}$, as $m_b V_0 \propto m_b r_c^2 \rho_c\propto\rho_c^{1/2}$.

The oscillation period of the ground-state soliton wavefunction is given by
\begin{eqnarray}
	\tau_{00}(\rho_c)&=& \frac{2\pi \hbar}{|E_{00}|} \simeq 39.9 \bigg(\frac{\rho_c}{\text{M}_\odot \text{pc}^{-3}}\bigg)^{-1/2} \text{ Myr},\;\;\;\;\; \label{eqn:NumGroundPeriod}
\end{eqnarray}
which expectedly agrees within $5 \%$ to the value observed in the simulated soliton wavefunction oscillations \cite{Schive:2019rrw} ($\tau_{00} = 116$ Myr and $\rho_c = 0.107$ M$_\odot$pc$^{-3}$). Irrespective of the presence of a background halo, the ground-state soliton both follows \eref{eqn:gammaDensityDist} in density distribution and appears well virialized. In physically well-motivated setups where the gravitational potential vanishes at infinity, this scaling relation is expected to hold as long as the halo contribution to the total potential is subdominant or marginally comparable to that of soliton. Granted that the solitonic core arises as the ground-state eigenfunction to the coupled SP equations, the de Broglie relation has been used to estimate the fluctuation timescale of FDM, giving $2\pi\hbar/m_b \sigma_\text{FDM}^2 \sim \tau_{00}$ (e.g. \cite{Marsh:2018zyw, Hui:2020hbq}), where $\sigma_\text{FDM}$ denotes the velocity dispersion of FDM. Note that however $\tau_{00}$ is not a gauge-invariant physical observable. Even more crucially, the true soliton oscillation timescale has $\tau_\text{soliton} \simeq 2.3 \tau_{00}$, as detailed below.

\begin{figure}
	\includegraphics[width=\linewidth]{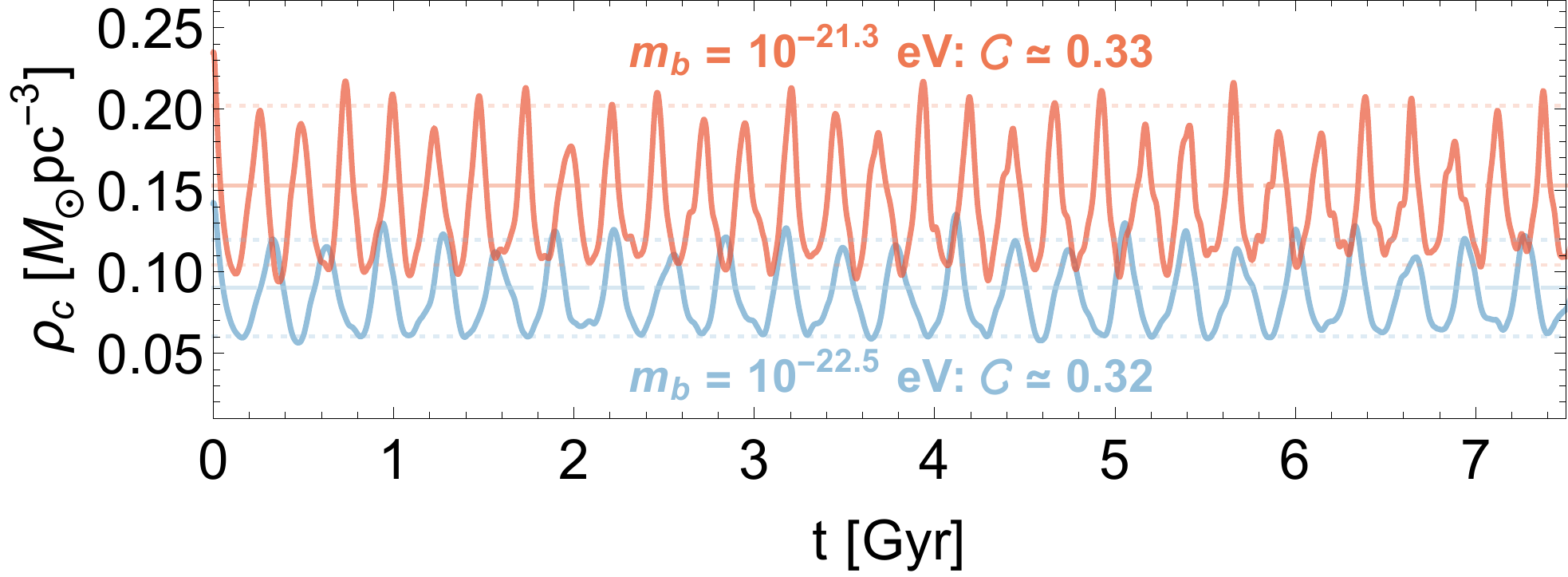}
	\includegraphics[width=\linewidth]{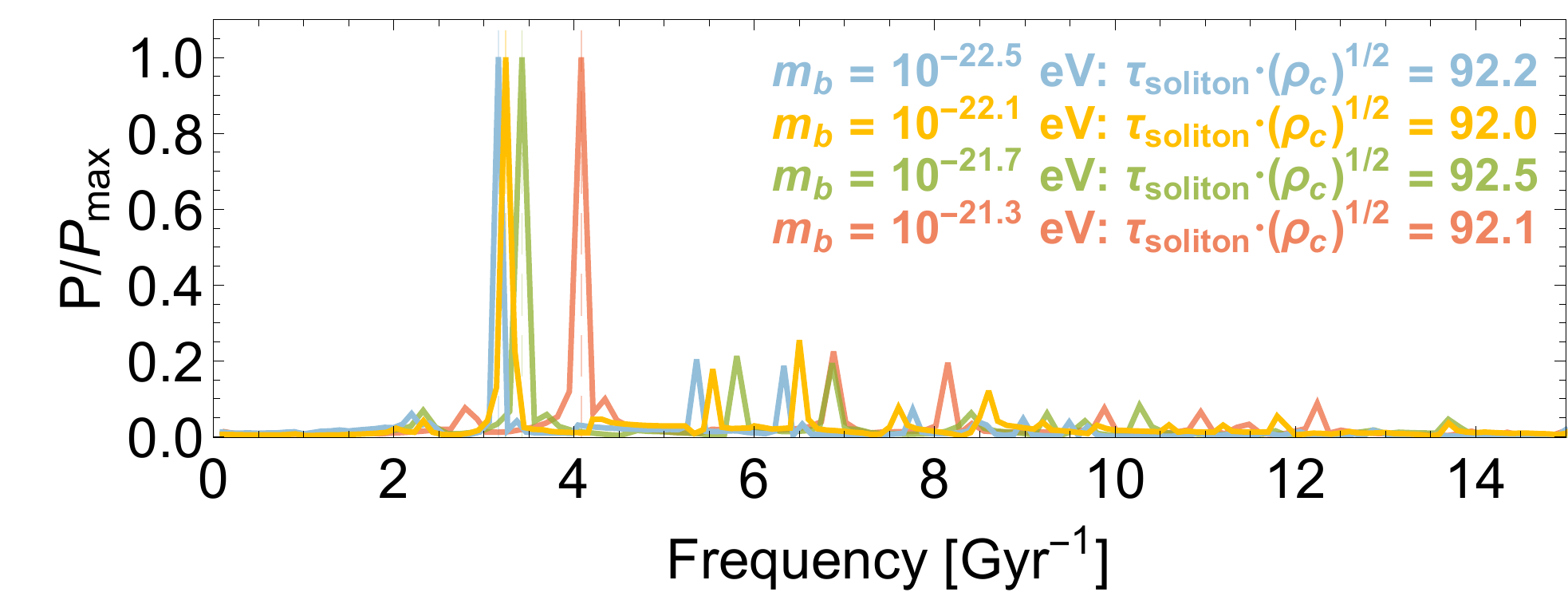}
	\caption{\textit{Top:} Peak core density $\rho_c$ in a perturbed soliton for $m_b=10^{-22.5}$ (blue) and $10^{-21.3}$ eV (red). The time-averaged densities and extrema (determined by $\mathcal{C}$) are denoted by dashed and dotted lines respectively. \textit{Bottom:} Power spectra of the density fluctuations for $m_b = 10^{-22.5}$ (blue), $10^{-22.1}$ (yellow), $10^{-21.7}$ (green), and $10^{-21.3}$ eV (red). The period of soliton density oscillations depends only on the peak core density, scaling as $\tau_\text{soliton} \propto \rho_c^{-1/2}$ (\eref{eqn:NumFirstExcitedPeriod}).}
	\label{fig:Gamer2}
\end{figure}

We simulate the evolution of a perturbed soliton using the code \texttt{GAMER} \cite{Schive:2017sdo}, where the initial density profiles of six isolated solitons with $m_b = 10^{-22.5}$\textendash$10^{-20.5}$ eV are artificially perturbed and then evolved for $\sim 8$\textendash10 Gyr (top panel of Fig \ref{fig:Gamer2}). For a given unperturbed soliton, we fix $\rho_c$ (adopted from Fig. \ref{fig:SolitonMass}) and slightly alter the outer slope of the soliton profile (e.g. changing the power in \eref{eqn:gammaDensityDist} from $-8$ to $-8.5$). The perturbed soliton then evolves till the oscillations stabilize, after which we adopt the time-averaged peak density as the genuine $\rho_c$ of this oscillating soliton. The soliton retains spherical symmetry during evolution. Our simulations show that the solitary density oscillations exhibit a characteristic timescale
\begin{eqnarray}
	\tau_\text{soliton}(\rho_c) \simeq 92.1 \bigg(\frac{\rho_c}{\text{M}_\odot \text{pc}^{-3}}\bigg)^{-1/2} \text{ Myr}, \label{eqn:NumFirstExcitedPeriod}
\end{eqnarray}
with a chi-square $\chi^2 = 3.79\times 10^{-3}$ by fitting the frequency spectra (bottom panel of Fig. \ref{fig:Gamer2}). The result is independent of both $m_b$ and the oscillation amplitude $\mathcal{C}$; $\tau_\text{soliton}\rho_c^{1/2}$ varies less than $\mathcal{O}(10^{-2})$ across the range $\mathcal{C} = 0.08$\textendash$0.35$ for different $m_b$ in our simulations. This genuine dynamical timescale of soliton density oscillations is universal, agreeing within 1\% to the values observed in fully self-consistent simulations: soliton-halo systems ($\mathcal{C} \simeq 0.33$) \cite{Veltmaat:2018dfz} and a stabilized soliton system after experiencing significant halo stripping ($\mathcal{C} \simeq 0.25$) \cite{Schive:2019rrw} (see also Sec. \ref{subsec:sim_fdm}). Note that the oscillation amplitude can nevertheless exhibit small variations.

To obtain an analytical estimate of $\tau_\text{soliton}$, we work perturbatively by considering $\mathcal{C} \rightarrow 0$. To the first-order analysis presented in Appendix \ref{app:PerturbationAppendix}, the coupled perturbations in the wavefunction and sourcing potential are related via \pCAeref{eqn:appRadialPerturbationEq}\pCCeref{app:CoupledDensity}. The eigenfunctions for a unperturbed soliton derived in Appendix \ref{app:SolitonEnergyLevels} conveniently offer a complete set of orthonormal basis to decompose the perturbed wavefunction $\delta u(\gamma, t)$. The resulting order-of-magnitude estimate \eref{eqn:PerturbationPeriod} yields $\tau_p(\rho_c) \simeq 0.58 \tau_\text{soliton}(\rho_c)$; whether high-order contributions can resolve the $\sim 40\%$ discrepancy between $\tau_p(\rho_c)$ and $\tau_\text{soliton}(\rho_c)$ is left to future work. In parallel, the interpretation proposed by Li et al. \cite{Li:2020ryg} attributes $\tau_\text{soliton}$ to the prominent interference timescale between ground-state and first-excited-state eigenfunctions. The reconstructed timescale, $\tau_\text{rec} \simeq 0.83 \tau_\text{soliton}$, depends on the energy difference $|E_{00}-E_{10}|$, which can also be directly read off from Table \ref{tab:SolitonTotalEnergy}.

Importantly, the timescales of the ground-state soliton wavefunction \eref{eqn:NumGroundPeriod} and soliton density oscillations \eref{eqn:NumFirstExcitedPeriod} are distinct, the ratio of which is a constant $\tau_\text{soliton}/\tau_{00} \simeq 2.3$ independent of $\rho_c, m_b,$ and $r_c$. This subtlety has not been correctly taken into account by the previous studies on the gravitational heating of a SC by an oscillating soliton \cite{Marsh:2018zyw}. We will show in the succeeding sections that SC heating due to soliton density oscillations is negligible irrespective of $m_b$, given that $\tau_\text{soliton}/\tau_\text{SC}$ deviates noticeably from unity.

\subsection{Constraints from Eridanus II}\label{sec:AstroConstraints}

The half-light radius of Eri II is estimated to be $r_\text{EII}=277\pm14$ pc \cite{Crnojevi:2016ApJ}, within which the mass enclosed is $M_\text{EII}^{\le r_\text{EII}}=1.2^{+0.4}_{-0.3}\times10^7$ M$_\odot$, corresponding to an average total mass density $\rho_\text{EII} \sim 0.135$ M$_\odot$pc$^{-3}$ \cite{Li:2016utv}. The inferred mass-to-light ratio indicates Eri II is DM-dominated \cite{Li:2016utv}. The central SC of Eri II appears to be an intermediate-age population ($\sim 3$\textendash$13$ Gyr) with a half-light radius of $r_\text{SC}=13\pm1$ pc \cite{Crnojevi:2016ApJ, Simon:2020qsf} and a mass of $M_\text{SC}^{\le r_\text{SC}} \sim 2000$ M$_\odot$ (assuming that the stellar mass-to-light ratio is unity) \cite{Li:2016utv}. The corresponding stellar mass density is $\rho_\text{SC} = 0.217$ M$_\odot$pc$^{-3}$.

\begin{figure}
	\centering
	\includegraphics[width=\linewidth]{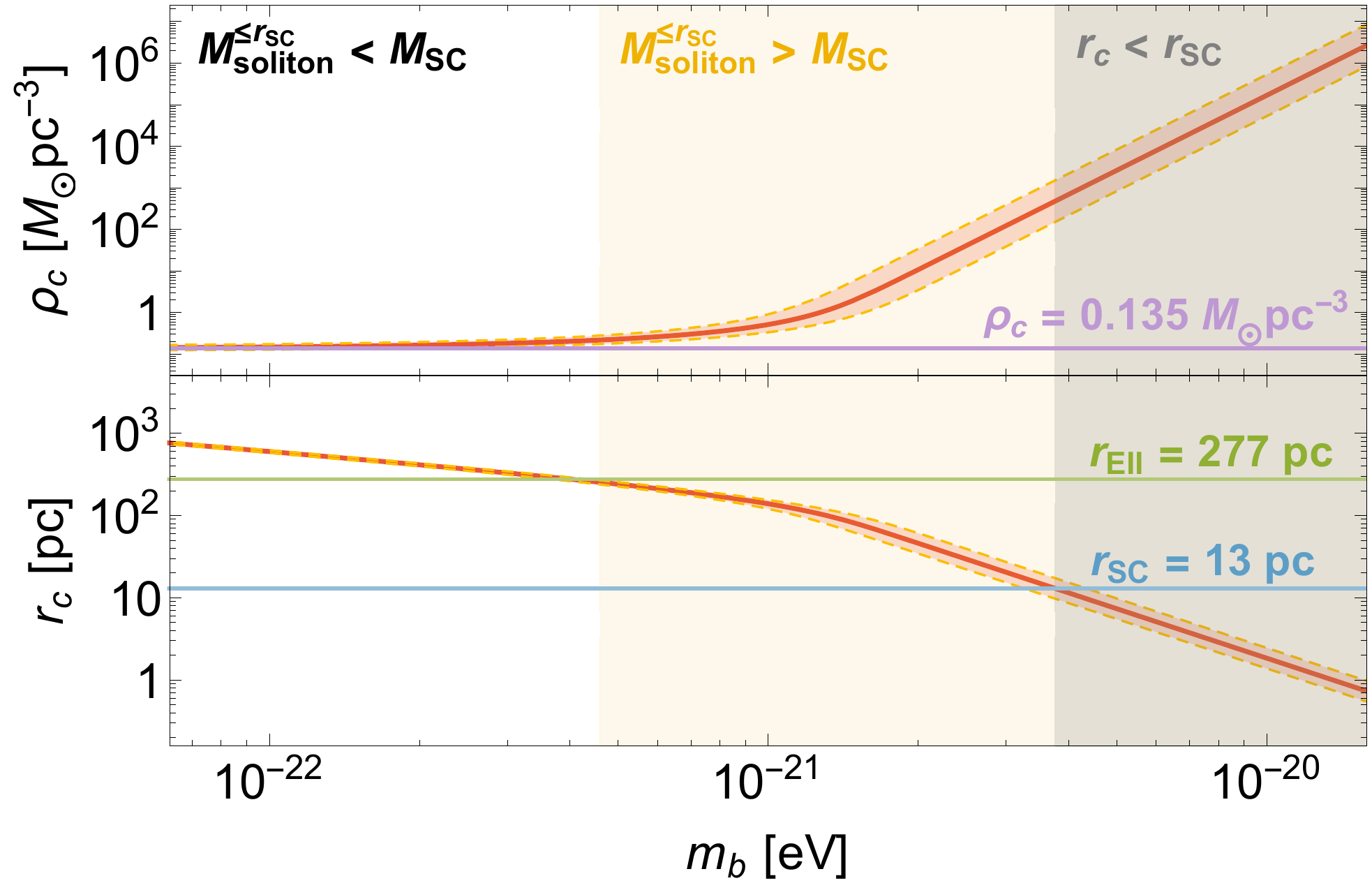}
	\caption{Values of $\rho_c$ (top panel) and $r_c$ (bottom panel) yielding $M_\text{soliton}^{\leq r_\text{EII}}= 1.2^{+0.4}_{-0.3}\times10^7$ M$_\odot$ at 68\% CL. (red shaded). The yellow shaded regions indicate where the gravitational potential within $r_\text{SC}$ is dominated by the soliton. The possible mass contribution from a background halo is excluded in this analysis, and thus the constraints on $\rho_c$ and $r_c$ as plotted here are most robust for $m_b \leq 1.5\times 10^{-21}$ eV, where $3.3 r_c \geq r_\text{EII}$ and consequently $M_\text{soliton}^{\leq r_\text{EII}}/M_\text{halo}^{\leq r_\text{EII}} \gg 1$. A background halo may be required to remedy the unphysically concentrated mass distribution within $r_\text{SC}$ in the gray shaded areas, where $r_c < r_\text{SC}$.}
	\label{fig:SolitonMass}
\end{figure}
The measurements of $M_\text{EII}^{\le r_\text{EII}}=1.2^{+0.4}_{-0.3}\times10^7$ M$_\odot$ place constraints on both $\rho_c$ and $r_c$, as plotted in Fig. \ref{fig:SolitonMass}; the soliton density profile thus reduces to an one-parameter scaling, uniquely determined by $m_b$ (or equivalently $\rho_c$). Note that the mass contribution for $\gamma \lesssim 3.3$ is completely dominated by the soliton \cite{Schive:2014dra, Mocz:2017wlg, Pozo:2020ukk}. We hence exclude the mass contribution of a possible background halo, the modeling of which is irrelevant in this analysis for $m_b\leq 1.5\times10^{-21}$ eV as $3.3r_c \geq r_\text{EII}$.

The range of $m_b$ shown in Fig. \ref{fig:SolitonMass} can be partitioned into four regions: (1) For $m_b \lesssim 4.6\times10^{-22}$ eV, $r_c > r_\text{EII}$ and the SC is gravitationally self-bound. The peak core density $\rho_c$ asymptotically approaches $\rho_\text{EII}$; $\rho_c \sim $ const then implies $r_c \propto m_b^{-1/2}$. (2) For $4.6\times 10^{-22}\eV \lesssim m_b \lesssim 1.5\times 10^{-21}$ eV, $r_\text{EII}/3.3 < r_c \leq r_\text{EII}$ and $\rho_c$ increases with $m_b$ as the soliton starts entering $r_\text{EII}$. As a result, the SC no longer remains self-bound (yellow shaded regions). (3) For $1.5\times 10^{-21}\eV \lesssim m_b \lesssim 3.8\times 10^{-21}$ eV, $r_\text{SC} < r_c \leq r_\text{EII}/3.3$ and the soliton is well within Eri II. As $M_\text{soliton} \simeq$ const, $\rho_c \propto m_b^6$ and $r_c \propto m_b^{-2}$. (4) Lastly for $m_b \gtrsim 3.8\times 10^{-21}$ eV, $r_c < r_\text{SC}$, indicating a soliton of mass $M_\text{EII}^{\leq r_\text{EII}}$ is contained within $r_\text{SC}$.

The FDM density typically drops by 1.5 to 3 orders of magnitude from $\rho_c$ at the soliton-halo profile transition for a wide range of possible distributions \cite{Schive:2014hza, Pozo:2020ukk, Mocz:2017wlg}. A conservative estimate suggests $M_\text{halo}^{\leq r_\text{EII}}/M_\text{soliton}^{\leq r_\text{EII}} \gtrsim \mathcal{O}(1)$ for $r_c \lesssim r_\text{EII}/4$, which implies the mass contribution of a background halo is still subordinate, albeit possibly not negligible for $1.5\times10^{-21} \lesssim m_b \lesssim 3.8\times10^{-21}$ eV. For $m_b \gtrsim 3.8\times 10^{-21}$ eV (gray shaded regions in Fig. \ref{fig:SolitonMass}), the soliton-only constraints become likely unphysical as the mass of entire Eri II is mostly contained within $r_\text{SC}$, which signals the presence of a background halo to prevent $r_c$ from shrinking below $r_\text{SC}$. Once the halo is introduced, however, the SC could be vulnerable to tidal disruption due to soliton random walk excursions \cite{Schive:2019rrw} or diffusion heating due to granular density fluctuations in the halo \cite{Marsh:2018zyw, Amr2020}. A more thorough numerical investigation is needed nevertheless, as \cite{Schive:2019rrw} only tested the impact of soliton random walk for $m_b = 8\times 10^{-23}\eV$ and the efficiency of diffusion heating has not been rigorously confirmed.

\section{Star Cluster Heating}\label{Heating}
\begin{figure}
	\includegraphics[width=\linewidth]{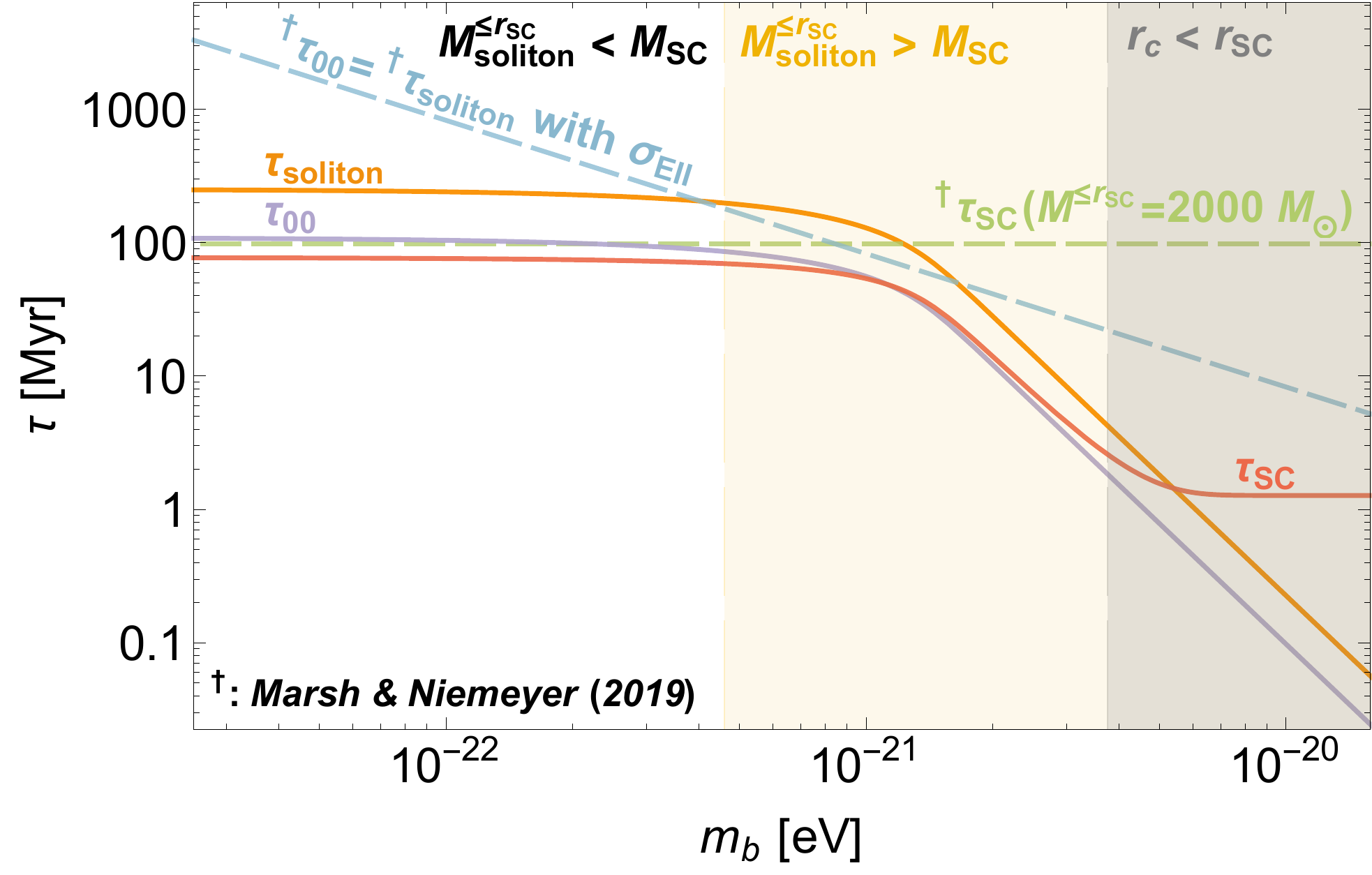}
	\caption{Dynamical timescales of Eri II star cluster (SC) $\tau_\text{SC}$ (\eref{eqn:MassOrbitalPeriod}, red), soliton density fluctuations $\tau_\text{soliton}$ (\eref{eqn:NumFirstExcitedPeriod}, orange), and the ground-state soliton wavefunction $\tau_{00}$ (\eref{eqn:NumGroundPeriod}, purple) in our analysis (solid). The halo mass contribution is excluded (Fig. \ref{fig:SolitonMass}). The apparent discrepancy between the aforementioned timescales (solid) and those adopted by \MN\,\cite{Marsh:2018zyw} (dashed, $^\dagger$) are discussed in Sec. \ref{app:TragetPaper}.}
	\label{fig:Gamer2PLOT2}
\end{figure}
The survival of Eri II SC was previously investigated by \MN\, to constrain the viable FDM particle mass. They considered the effect of gravitational heating on the SC caused by soliton density oscillations and identified several resonance bands with efficient heating between $10^{-21} \text{ eV}\lesssim m_b \lesssim 5\times 10^{-21}$ eV. On the contrary, we demonstrate in this section that, with an improved estimate of $\tau_{\rm soliton}$, SC heating becomes negligible for $m_b \lesssim 3.8\times10^{-21}\eV$.

We model the single-mode soliton density oscillations with an amplitude $\mathcal{C}$, frequency $\omega_\text{osc}=2\pi/\tau_\text{osc}$, and an arbitrary phase $\phi = \left[0,2\pi\right)$ by
\begin{eqnarray}
\rho_c \Big[1+ \mathcal{C}\sin(\omega_\text{osc}t+\phi)\Big]\eee\rho_cs(t), \label{eqn:SolitonOsc}
\end{eqnarray}
which propagates the time-dependence to the soliton radius $r_c s(t)^{-1/4}$ (\eref{eqn:SolitonRadiusEriII}) and consequently the soliton enclosed mass $M_\text{soliton}^{\leq r}(\gamma = r r_c^{-1} s(t)^{1/4}; \rho_c s(t), m_b)$ (\eref{eqn:SolitonEncMass}). Here both $\rho_c$ and $r_c$ are the unperturbed (time-averaged) values; the core is assumed to retain the soliton profile during oscillations. We treat $\omega_\text{osc}$ as a free variable to locate the resonance bands.

\begin{figure}
	\includegraphics[width=\linewidth]{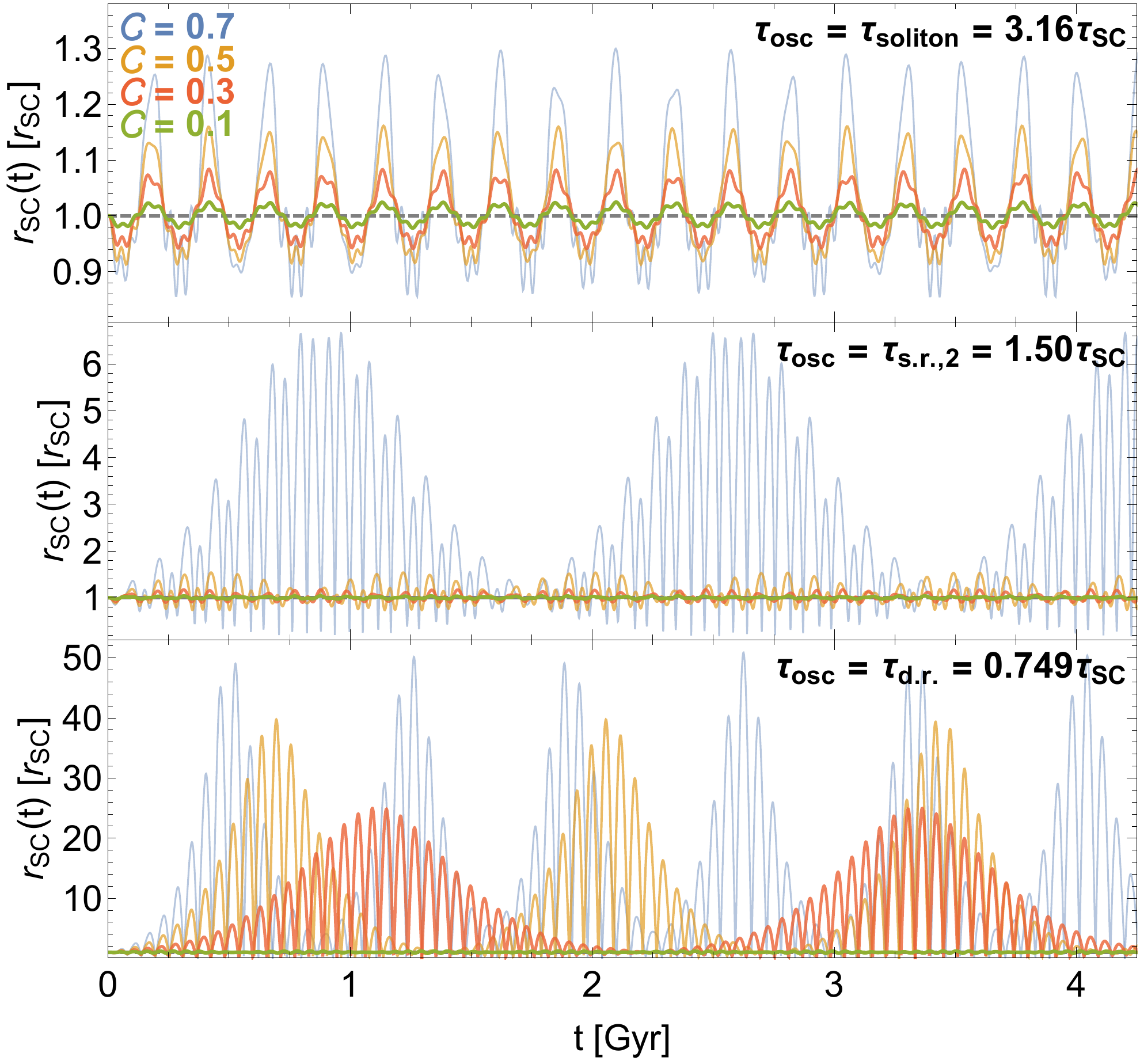}
	\caption{Time evolution of test star's orbital radius normalized to $r_\text{SC} = 13$ pc, for $\mathcal{C} = $ 0.1 (green), 0.3 (red), 0.5 (yellow), and 0.7 (blue), fixing $m_b = 10^{-22}$ eV. The top, middle, bottom panels has $\tau_\text{osc} = \tau_\text{soliton}$, $\tau_\text{s.r.,2}$, and $\tau_\text{d.r.}$ (\pAeref{eqn:DirectResonanceaeff}\pBeref{eqn:Superharmonics}) respectively; \eref{eqn:MassOrbitalPeriod} yields $\tau_\text{SC} = 75.8$ Myr. Gravitational heating of Eri II SC is ineffective, suggested by the realistic modeling of soliton oscillations with $\tau_\text{osc} = \tau_\text{soliton}$ (see also Sec. \ref{subsec:sim_fdm}).}
	\label{fig:TidalHeatingResultsC}
\end{figure}
Consider a star of mass $m_\star$ at an initial distance $r_0=r_\text{SC} = 13$ pc with respect to the center of mass of SC, following a similar setup in \MN\, (see also Sec. \ref{app:TragetPaper}). Here we assume the centers of the SC and Eri II coincide to a good approximation. The unperturbed circular orbit has an orbital frequency \footnote{The cored density profiles of both a soliton and the SC, roughly following $M^{\leq r} \propto r^3$ within their respective cutoff radii, suggest that this estimate of $\omega_{\text{SC},0}$ is insensitive to the choice of $r_\text{SC}$.}
\begin{eqnarray}	
	\omega_{\text{SC},0} &&=\sqrt{\frac{G\big(M_\text{SC}^{\le r_\text{SC}}+M_\text{soliton}^{\leq r_\text{SC}}\big)}{r_\text{SC}^3}}. \label{eqn:MassOrbitalPeriod}
\end{eqnarray}

Figure \ref{fig:Gamer2PLOT2} compares the relevant timescales of soliton dynamics under the soliton-only constraints (i.e. ignoring the mass contribution from an outer halo), where $\tau_\text{SC} = 2\pi/\omega_{\text{SC},0}$ (red), $\tau_\text{soliton}$ (orange), and $\tau_{00}$ (purple) are plotted in solid. For $m_b \lesssim 4.6\times 10^{-22}$ eV, the SC orbital period $\tau_\text{SC}$ roughly remains constant (so are $\tau_{00}$ and $\tau_\text{soliton}$) since $\rho_c \sim$ const. Moreover, $\tau_\text{soliton}/\tau_\text{SC} \sim 2$\textendash3 for $m_b \lesssim3.8\times 10^{-21}$ eV. For $m_b \gtrsim 3.8\times 10^{-21}$ eV, $\tau_\text{SC}$ also approaches a constant value as $r_c \lesssim r_\text{SC}$ and $M_\text{soliton}^{\leq r_\text{SC}}\sim M_\text{EII}^{\leq r_\text{EII}}$, leading to $\tau_\text{soliton}/\tau_\text{SC} \lesssim 1$. However if a background halo is introduced to ensure $r_c \geq r_\text{SC}$, then $\tau_\text{soliton}/\tau_\text{SC} \gtrsim 2$ even for $m_b \gtrsim 3.8\times10^{-21}$ eV. See Sec. \ref{app:TragetPaper} regarding remarks about $^\dagger\tau_{00} \eee {^\dagger\tau_\text{soliton}}$ (blue, dashed) and $^\dagger\tau_\text{SC}$ (green, dashed) presented in \MN, which are apparently inconsistent with our analysis.

\begin{figure}
	\includegraphics[width=\linewidth]{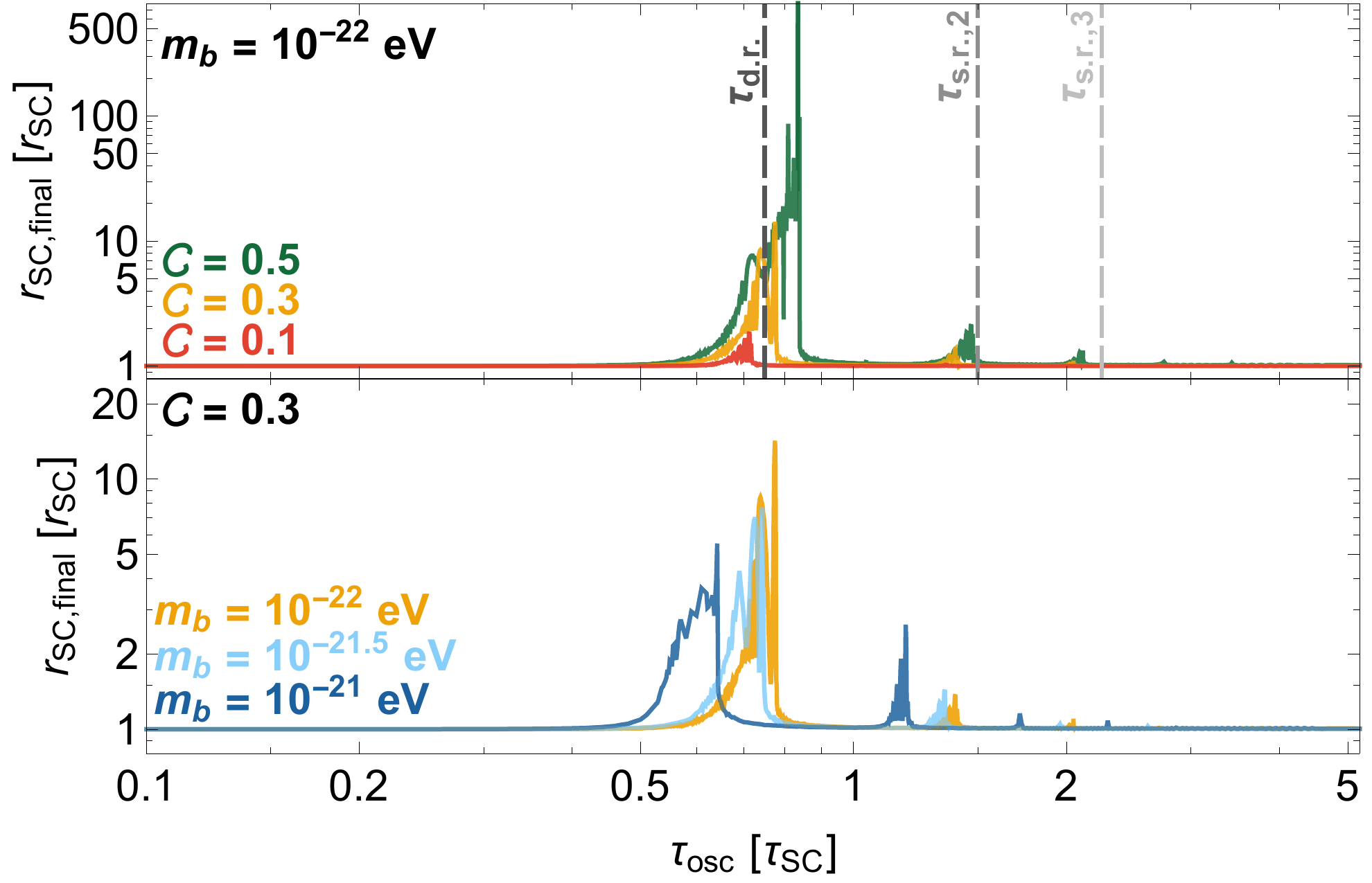}
	\caption{Time-averaged radial migration of a test star over 10 Gyr. \textit{Top:} We compare three fiducial values of $\mathcal{C} = 0.1$ (red), $0.3$ (yellow), and $0.5$ (green), fixing $m_b = 10^{-22}$ eV. Direct resonance $\tau_\text{d.r.}$ and superharmonic resonances $\tau_\text{s.r.}$ are given by \pAeref{eqn:DirectResonanceaeff}\pBeref{eqn:Superharmonics}. \textit{Bottom:} We take $m_b = 10^{-22}$ eV (yellow), $10^{-21.5}$ eV (light blue), and $10^{-21}$ eV (dark blue), fixing $\mathcal{C} = 0.3$. The efficiency of gravitational heating peaks at $\tau_\text{osc}/\tau_\text{SC} \simeq 0.5$\textendash$0.8$ and is negligibly low for the realistic case with $\tau_\text{osc}/\tau_\text{SC} \simeq 2\textendash3$.}
	\label{fig:NumericalRAverage}
\end{figure}
The time evolution of the orbital radius reads
\begin{eqnarray}
	\ddot{r} = \omega_{\text{SC}}^2 r- \frac{G\big[M_\text{SC}^{\le r_\text{SC}}+M_\text{soliton}^{\leq r(t)}(\gamma;\rho_cs(t),m_b)\big]}{r^2}, \;\;\;\;\;\;\;\label{eqn:TidalHeatingMass}
\end{eqnarray}
where the conservation of angular momentum gives $\omega_\text{SC}^2r = \omega_{\text{SC},0}^2r_\text{SC}^4r^{-3}$.The same radial equation of motion (\eref{eqn:appTidalHeatingFullyGeneral}) can also be obtained via the Hamiltonian formalism by treating the density fluctuations as a time-dependent perturbation in the ground-state soliton potential (see Appendix \ref{app:HamiltonianPerturbation}). We then linearize \eref{eqn:TidalHeatingMass} to the first non-vanishing order in $\gamma$ and identify two types of instabilities: the direct resonance $\tau_\text{d.r.}$ (\eref{eqn:DirectResonanceaeff}) and a series of secondary superharmonic resonances $\tau_{\text{s.r.},n} \simeq n \tau_\text{d.r.}$ for $n \geq 2, n \in \mathbb{Z}^{+}$ (\eref{eqn:Superharmonics}), as detailed in Appendix \ref{app:Resonance}.

Figure \ref{fig:TidalHeatingResultsC} compares the time evolution of the orbital radius obtained by numerically integrating \eref{eqn:TidalHeatingMass} for $\mathcal{C} = 0.1$ (green), $0.3$ (red), $0.5$ (yellow), and $0.7$ (blue), fixing $r_0 = r_\text{SC} = 13$ pc, $\phi =0$, and $m_b = 10^{-22}$ eV (corresponding to $\rho_c = 0.149$ M$_\odot$pc$^{-3}$). The top, middle, and bottom panels adopt $\tau_\text{osc} = \tau_\text{soliton}, \tau_\text{s.r.,2},$ and $\tau_\text{d.r.}$ respectively. The orbit is non-Keplerian for $\mathcal{C} \neq 0$. We observe that gravitational heating is inefficient for the realistic modeling of soliton oscillations with $\tau_\text{osc} = \tau_\text{soliton}$; this conclusion is further supported by the self-consistent numerical simulations elaborated in Sec. \ref{subsec:sim_fdm}. The effect of orbital resonances becomes more pronounced when $\tau_\text{osc}/\tau_\text{SC} \rightarrow 1$ falls on resonance bands and for large $\mathcal{C}$.

Figure \ref{fig:NumericalRAverage} shows the shift in orbital radius $\smash{r_\text{SC,final} \eee\int_{9 \text{ Gyr}}^{10 \text{ Gyr}}r(t) dt \big/1 \text{ Gyr}}$ \footnote{The resulting $r_\text{SC,final}$ converges for alternative choices of averaging over $8$\textendash$10$ Gyr or adopting all time maximum.}. The results converge for arbitrary choices of $\phi$ and $ -25\leq \dot{r}_0\leq 25 $. The top panel plots the time-averaged radial migration for $\mathcal{C} = 0.1$ (red), $0.3$ (yellow), and $0.5$ (green) as a function of $\tau_\text{osc}/\tau_\text{SC}$ with $m_b = 10^{-22}$ eV. In the bottom panel, we take $m_b = 10^{-22}$ eV (yellow), $10^{-21.5}$ eV (light blue), and $10^{-21}$ eV (dark blue), fixing $\mathcal{C} = 0.3$. Non-trivial orbital resonances are observed only for $\tau_\text{osc}/\tau_\text{SC} \simeq 0.5$\textendash$0.8$, which confirms a negligible heating effect of SC given $\tau_\text{soliton}/\tau_\text{SC} \sim 2$. The resonance bands widen as $\mathcal{C}$ increases. For large $m_b$, the constrained $\rho_c$ and consequently the (initial) angular momentum increase, which qualitatively shifts both $\tau_\text{d.r.}$ and $\tau_\text{s.r.,n}$ towards smaller $\tau_\text{osc}/\tau_\text{SC}$.

\begin{figure}
	\includegraphics[width=\linewidth]{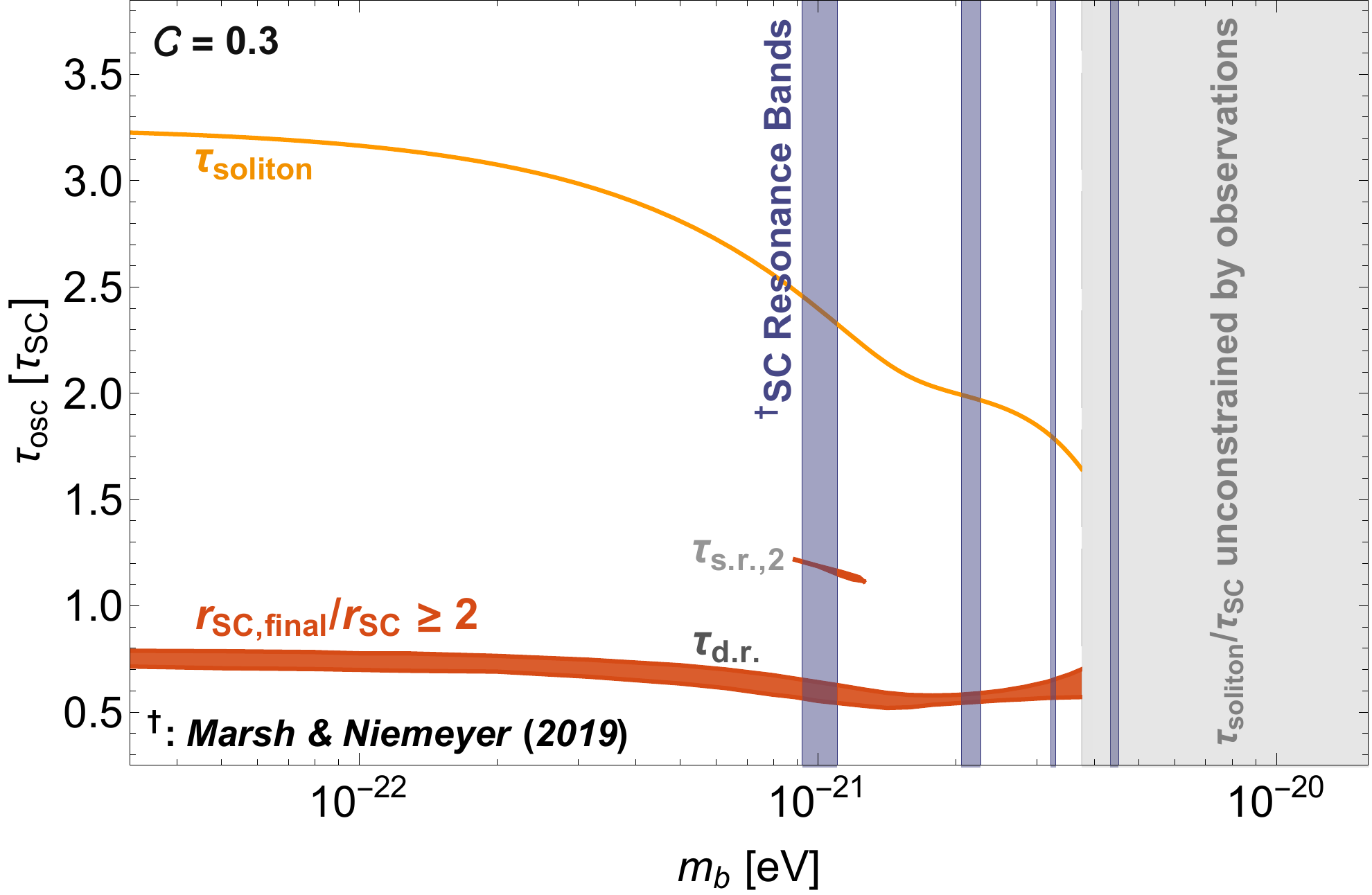}
	\caption{Stability diagram of \eref{eqn:TidalHeatingMass}, fixing $\mathcal{C} = 0.3$ with soliton-only constrained $\rho_c$ and $r_c$ applicable for $m_b \lesssim 4\times 10^{-21}$ eV (Fig. \ref{fig:SolitonMass}). The soliton oscillation timescale $\tau_\text{soliton}$ (orange) shows no overlap with $\tau_\text{d.r.}$ or $\tau_{\text{s.r.},2}$ (red shaded regions where $r_\text{SC,final}/r_\text{SC} \geq 2$ over $10$ Gyr). The ineffective SC heating is in clear contradiction with the result by \MN\, \cite{Marsh:2018zyw} (purple resonance bands); see also Sec. \ref{app:Comment2}.}
	\label{fig:TidalHeatingResultsRhoc}
\end{figure}
Figure \ref{fig:TidalHeatingResultsRhoc} plots the SC stability diagram; the red shaded areas indicate the parameter space with efficient orbital heating, defined as $r_\text{SC,final}/r_\text{SC} \geq 2$ over 10 Gyr. The genuine oscillation timescale of an FDM soliton $\tau_\text{soliton}$ (orange) does not intersect with either $\tau_\text{d.r.}$ or $\tau_{\text{s.r.},2}$; the heating mechanism is therefore inefficient. The SC resonance bands identified in NM2019 are shown in purple; this stark inconsistency is further discussed in Sec. \ref{app:TragetPaper}. For $m_b \gtrsim 4\times 10^{-21}$ eV (gray shaded), the significant uncertainty in subhalo profile modeling propagates to the soliton-only constraints on $\rho_c$ and $r_c$, rendering $\tau_\text{soliton}/\tau_\text{SC}$ poorly constrained by observations.

In summary, the single-test-star toy model firmly indicates that gravitational heating from an oscillating soliton is inefficient. For $m_b \lesssim 4.6\times 10^{-22}$ eV, the SC is self-bound as $\rho_c \sim$ const and $\rho_c\lesssim \rho_\text{SC} = 0.217$ M$_\odot$pc$^{-3}$. The $m_b$-independent relation in the dynamical timescales, $\tau_\text{soliton} > 2\tau_\text{SC}$, results in a negligible effect of SC heating. For $m_b \gtrsim 4.6\times 10^{-22}$ eV, we have $\rho_c \gtrsim \rho_\text{SC}$ as $\rho_c$ increases with $m_b$. The SC is not self-bound and consequently $\tau_\text{SC}$ is determined by $\rho_c$. The timescales $\tau_\text{SC}$ and $\tau_{00}$ differ $\lesssim 30\%$ under the soliton-only constraints for the applicable range of $m_b$, as detailed in Sec. \ref{sec:SolitonTimeScale}. Consequently, we have $\tau_\text{SC} \simeq \tau_{00} \sim \tau_\text{soliton}/2$ and hence $\tau_\text{soliton}/\tau_\text{SC} \simeq 2$, suggesting again ineffective heating.

\begin{figure}
	\includegraphics[width=\linewidth]{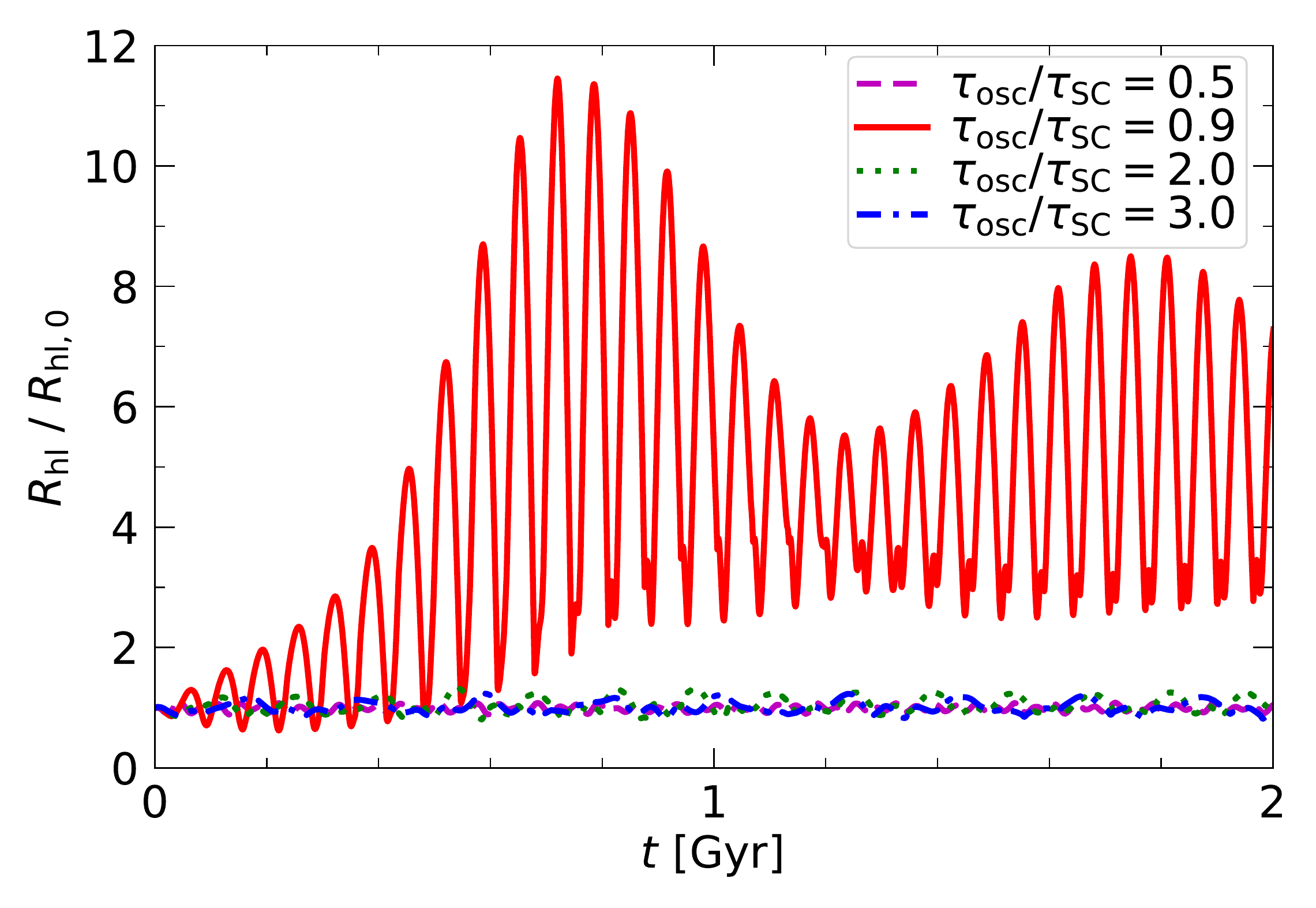}
	\caption{Time evolution of the SC's projected half-light radius $\Rhl$ in an oscillating external soliton potential with period $\tosc$. $\Rhl$ is normalized to its initial value $\Rhli$ and $\tosc$ to the characteristic timescale of the SC $\tSC$. Among the four representative cases with different $\tosc/\tSC$, only the case $\tosc/\tSC=0.9$ exhibits significant heating, with $\Rhl$ increasing by an order of magnitude within $1\Gyr$. See also Figs. \ref{fig:simu_ext_proj_dens} and \ref{fig:simu_ext_profile}.}
	\label{fig:simu_ext_Rhl_evolution}
\end{figure}

\section{Simulations}
\label{sec:sim}

In this section, we perform three-dimensional simulations to compare against
the single-particle toy model presented in Sec. \ref{Heating}. We first model
the soliton oscillations as an external potential, followed by self-consistent
FDM simulations.

\subsection{N-body Simulations with an External Potential}
\label{subsec:sim_ext}

\subsubsection{Simulation setup}
\label{subsubsec:sim_ext_setup}

\begin{figure}
	\includegraphics[width=\linewidth]{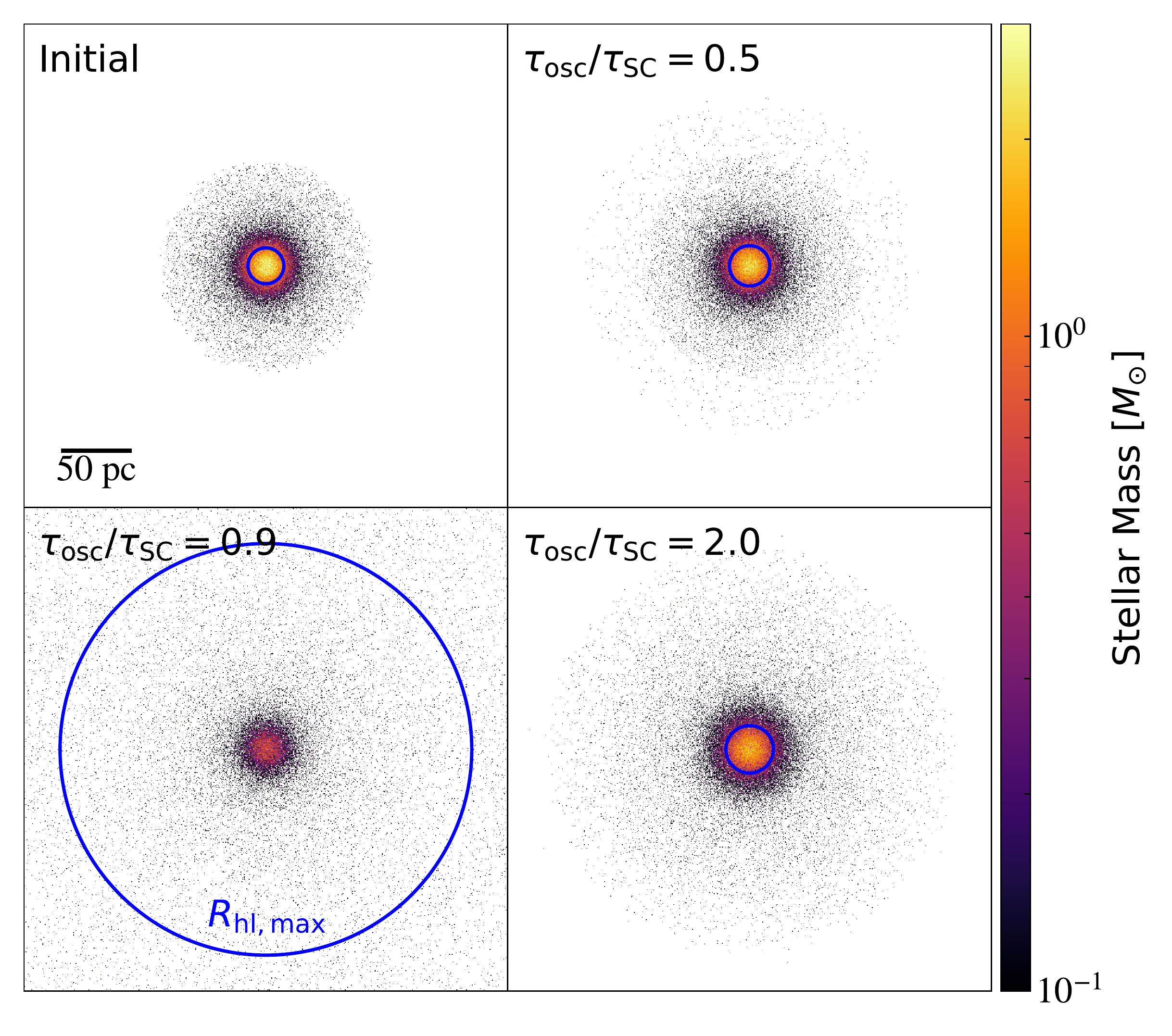}
	\caption{Projected mass of the SCs and the associated half-light radii (blue circles). The top left panel shows the initial condition. The other three show the results with different values of $\tosc/\tSC$ at the respective epoch with a maximum half-light radius $\Rhlmax$ (see \fref{fig:simu_ext_Rhl_evolution}). Heating by an oscillating soliton is only manifest for $\tosc/\tSC=0.9$. Visualization is done with \texttt{yt} \cite{yt}.}
	\label{fig:simu_ext_proj_dens}
\end{figure}

The SC is modeled by a Plummer sphere with a total mass $8.7\times10^3\Msun$ (within a cut-off radius of $100\pc$), a projected half-light radius $\Rhli = 22\pc$, and a corresponding peak stellar mass density $\rhosc \sim 0.22\Msun\pc^{-3}$. We follow the treatment in \cite{Aarseth1974AA37183A} to sample the stellar velocity. The soliton is modeled by an external potential following \eref{eqn:VExcitedStatesAnalysis} that fixes $m_b=10^{-22} \eV$ and $r_c=650\pc$, yielding a soliton mass density $\rhosol \sim 0.11\Msun\pc^{-3}$. The centers of the SC and the soliton coincide.

The SC is self-bound but the gravitational influence of the soliton is non-negligible, as $\rhosc \sim 2 \rhosol$. We thus first evolve the SC in a \emph{fixed} soliton potential until it relaxes and then take this relaxed SC as the initial condition, which has a smaller projected half-light radius $\Rhli \sim 13\pc$ and a characteristic timescale $\tSC \sim 71\Myr$ (\eref{eqn:MassOrbitalPeriod}). The soliton then oscillates according to \eref{eqn:SolitonOsc} with varying $\Wosc=2\pi/\tosc$ for a fixed $\mathcal{C}=0.5$ over $2\Gyr$.

We use a direct N-body method with a fourth-order Hermite scheme, individual time-step, and GPU acceleration \cite{Schive:2007mn} to evolve the SC. The gravitational softening length is set to $0.1 \textrm{--} 1.0\pc$. The total number of particles is $2\times10^4 \textrm{--} 8\times10^4$, leading to a particle mass resolution of $\sim 0.4 \textrm{--} 0.1\Msun$. We have verified that the simulation results are insensitive to the adopted softening length and particle mass resolution. The maximum relative error in angular momentum conservation is less than $0.2\%$ in all simulations.

\subsubsection{Orbital evolution and heating efficiency}
\label{subsubsec:sim_ext_results}

\begin{figure}
	\includegraphics[width=\linewidth]{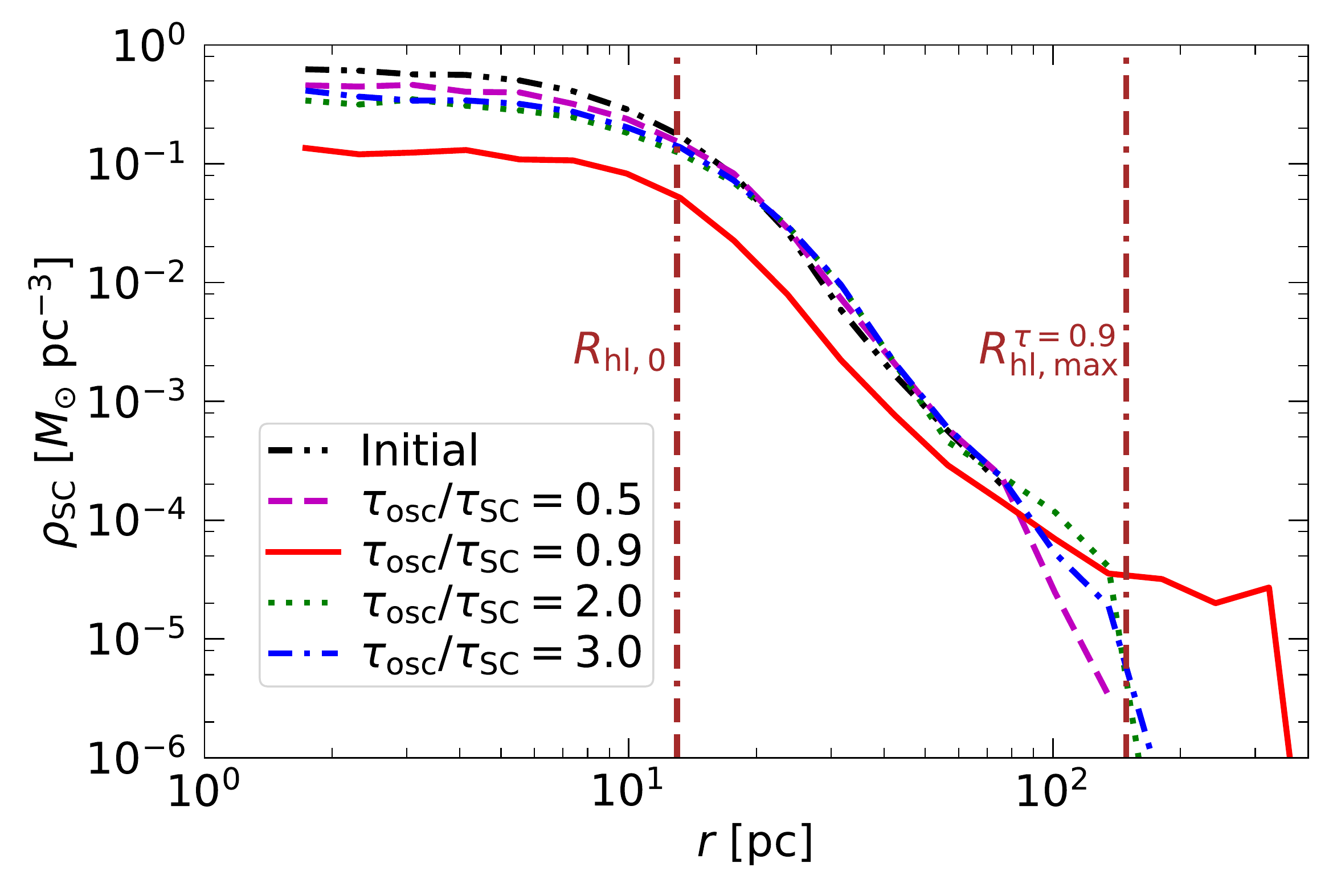}
	\caption{Stellar density profiles. We plot the initial condition (black) and the configurations at the respective epoch with maximum half-light radii for $\tosc/\tSC = 0.5$ (purple), $0.9$ (red), $2.0$ (green), and $3.0$ (blue) (see \fref{fig:simu_ext_Rhl_evolution}). The vertical lines mark the initial half-light radius ($\Rhli$) and the maximum half-light radius for $\tosc/\tSC=0.9$ ($\Rhlmax^{\tau=0.9}$), only where the effect of heating is appreciable.}
	\label{fig:simu_ext_profile}
\end{figure}

First, we present the results of four representative cases with $\tosc/\tSC= 0.5, 0.9, 2.0,$ and $3.0$. Figure \ref{fig:simu_ext_Rhl_evolution} shows the time evolution of the projected half-light radius $\Rhl(t)$. Only the case $\tosc/\tSC=0.9$ exhibits prominent heating, with a maximum half-light radius $\Rhlmax \sim 11.5 \Rhli$ at $t \sim 0.7\Gyr$. For other three fiducial values of $\tosc/\tSC$ deviating notably from unity, the SC is stable against an oscillating external soliton potential, with $\Rhlmax \lesssim 1.3 \Rhli$. Figure \ref{fig:simu_ext_proj_dens} compares the initial projected stellar mass with the resulting configurations when $\Rhl(t)=\Rhlmax$ for $\tosc/\tSC = 0.5, 0.9, 2.0$. The corresponding density profiles are plotted in \fref{fig:simu_ext_profile}. The efficiency of gravitational heating is greatly attenuated for $\tosc/\tSC \neq \mathcal{O}(0.9)$.

\begin{figure}
	\includegraphics[width=\linewidth]{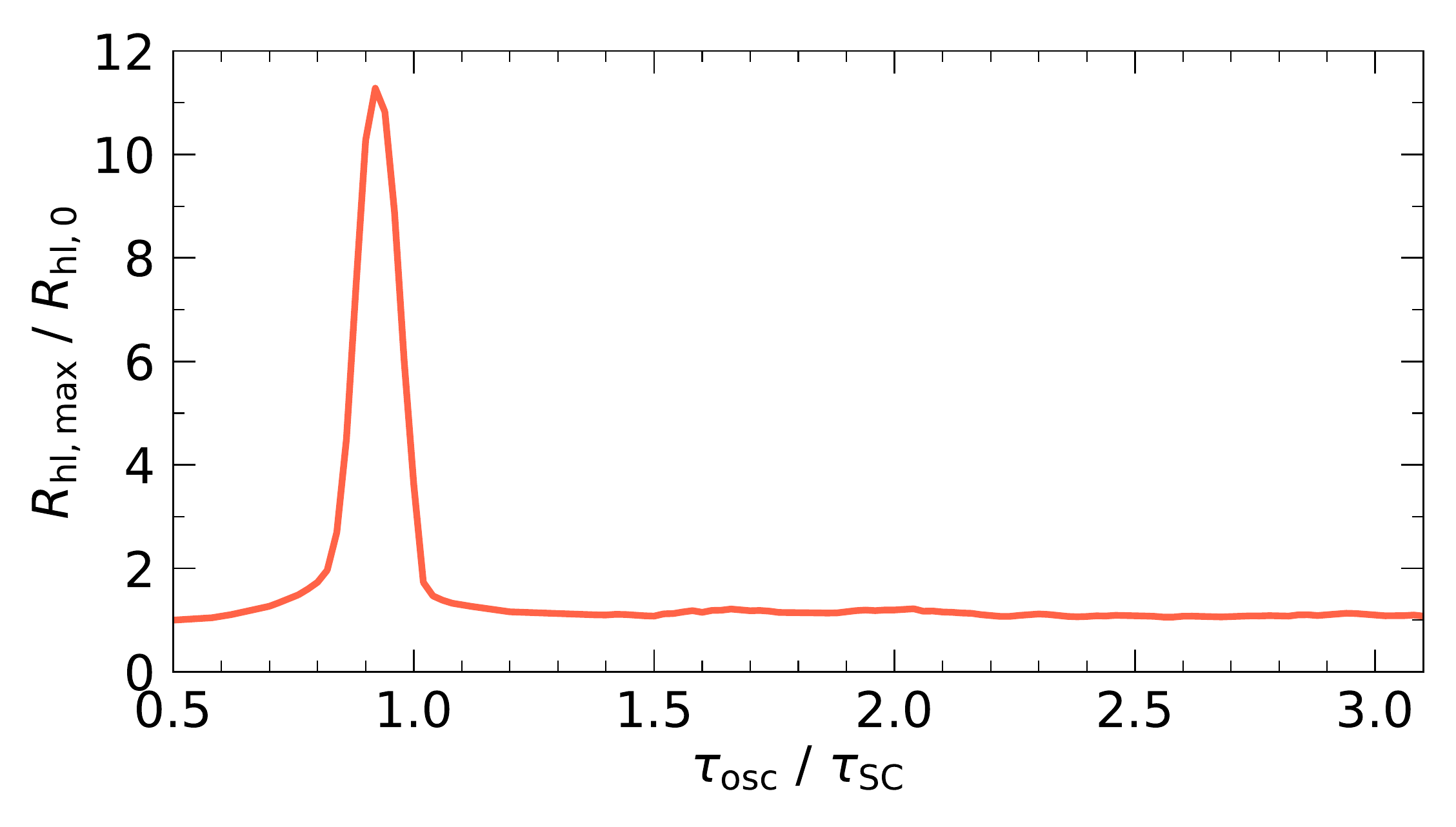}
	\caption{Maximum expansion of the SC's half-light radius caused by an external oscillating soliton potential with period $\tosc$. Gravitational heating is most prominent at $\tosc/\tSC \sim 0.9$ and becomes ineffective as $\tosc/\tSC$ deviates from unity.}
	\label{fig:simu_ext_Rhl_max}
\end{figure}

We then perform an extensive set of simulations to probe the heating efficiency with $0.5 \le \tosc/\tSC \le 3.1$. Figure \ref{fig:simu_ext_Rhl_max} shows $\Rhlmax$ as a function of $\tosc/\tSC$. The heating efficiency peaks around $\tosc/\tSC \sim 0.9$ and drops dramatically when $\tosc/\tSC$ deviates from unity. This result is in good agreement with the single-particle toy model (\fref{fig:NumericalRAverage}), where the heating efficiency peaks at $\tosc/\tSC = 0.835$ for $m_b = 10^{-22}$ eV and $\mathcal{C} = 0.5$. The half-mass containment for a Plummer profile lies between $0.483\Rhli$\textendash$1.52\Rhli$, within which the variation in $\tau_\text{SC}(r)$ is less than $\sim 50\%$ for $\rho_c/\rho_\text{SC} \geq 0.622$ consistent with observations. The assumption of no-shell crossing in the single-particle description is therefore largely applicable to the more realistic SC modeling.

\subsection{Self-consistent FDM Simulations}
\label{subsec:sim_fdm}

To further test whether the SC can survive within a realistic FDM soliton, in this subsection we evolve both the SC and soliton in a \emph{self-consistent} approach.

\subsubsection{Simulation setup}
\label{subsubsec:sim_fdm_setup}

The SC is embedded and evolved in the center of an FDM halo mimicking Eri II, following the simulation setup in \cite{Schive:2019rrw}. In this work, we focus on the case where the outer halo surrounding the soliton has been largely stripped away by an external tidal field; subsequently, soliton random walk excursions are greatly suppressed and the effect thereof can be safely ignored (see Figs. 3 and 4 therein). This assumption is crucial, for otherwise the off-center separation between the soliton and the SC can tidally disrupt the SC within $\sim 1\Gyr$, as demonstrated in \cite{Schive:2019rrw}.

\begin{figure}
	\includegraphics[width=\linewidth]{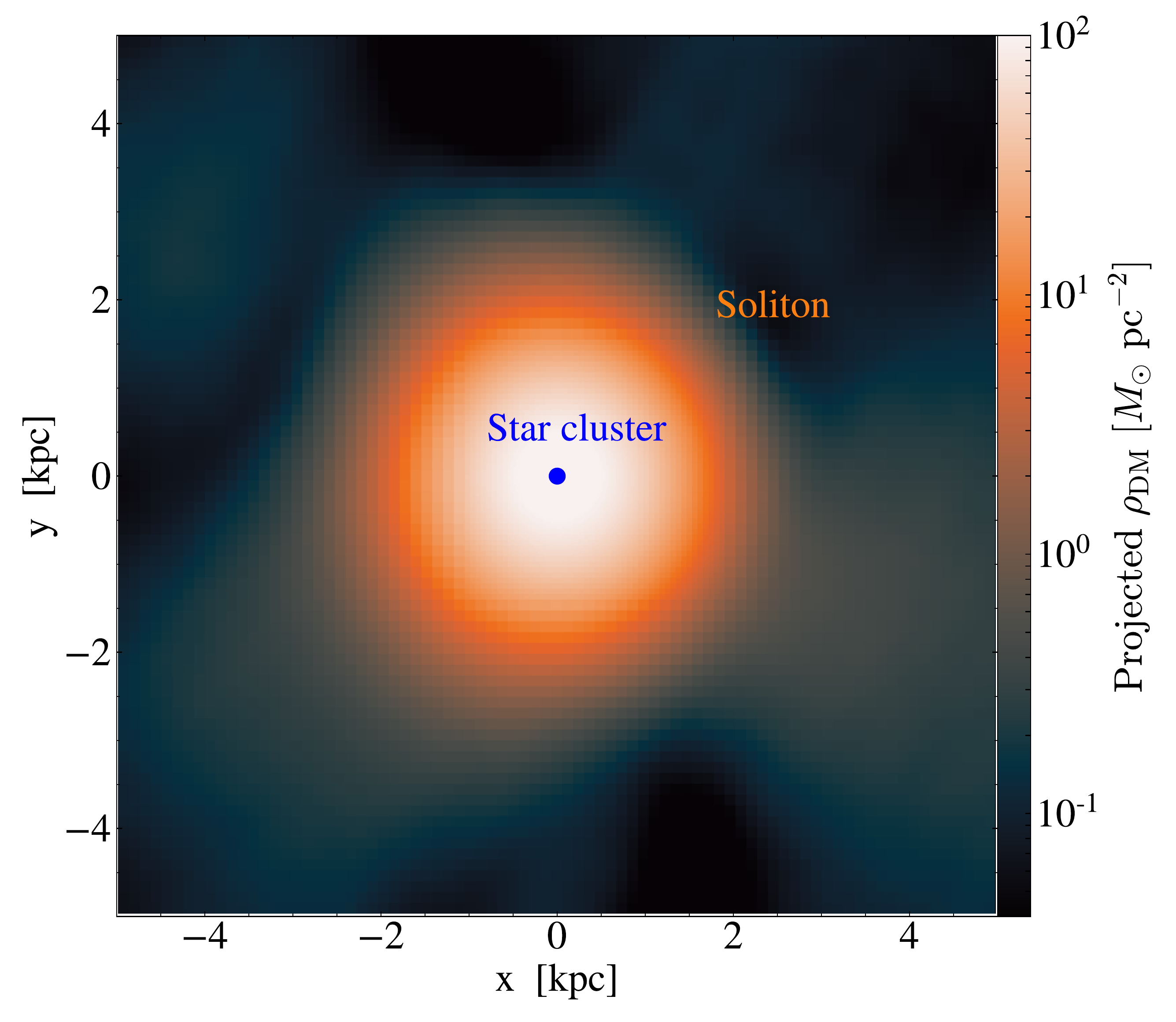}
	\caption{Projected dark matter density of the initial condition in a self-consistent FDM simulation. The bright region represents the soliton and the blue solid circle marks the SC. The ambient medium of soliton has been largely stripped away by an external tidal field. Visualization is done with \texttt{yt} \cite{yt}.}
	\label{fig:simu_dyn_proj_dens}
\end{figure}

The adopted FDM particle mass $m_b=8.0\times10^{-23} \eV$ and the soliton half-density radius $r_c \sim 830\pc$ together give $\rhosol \sim 6.5\times10^{-2}\Msun\pc^{-3}$. The SC parameters are the same as those adopted in the previous subsection, except that we do not need to relax the SC further due to a lower value of $\rhosol$ here. The corresponding characteristic timescale of the SC is $\tSC \sim 120\Myr$ (\eref{eqn:MassOrbitalPeriod}). The initial density distribution is illustrated in \fref{fig:simu_dyn_proj_dens}.

We use the code \texttt{GAMER} \cite{Schive:2017sdo} to evolve FDM and the SC self-consistently. It supports adaptive mesh refinement and hybrid CPU/GPU parallelization. We adopt a simulation box of size $250\kpc$ covered by a $128^3$ root grid and nine refinement levels, leading to a maximum spatial resolution of $\sim 3.8\pc$ capable of resolving the compact SC. The particle mass resolution is $3.6\times 10^{-2}\Msun$. See \cite{Schive:2019rrw} for more details. We run the simulations for $5\Gyr$.

\subsubsection{Heating efficiency}
\label{subsubsec:sim_fdm_results}

\begin{figure}
	\includegraphics[width=\linewidth]{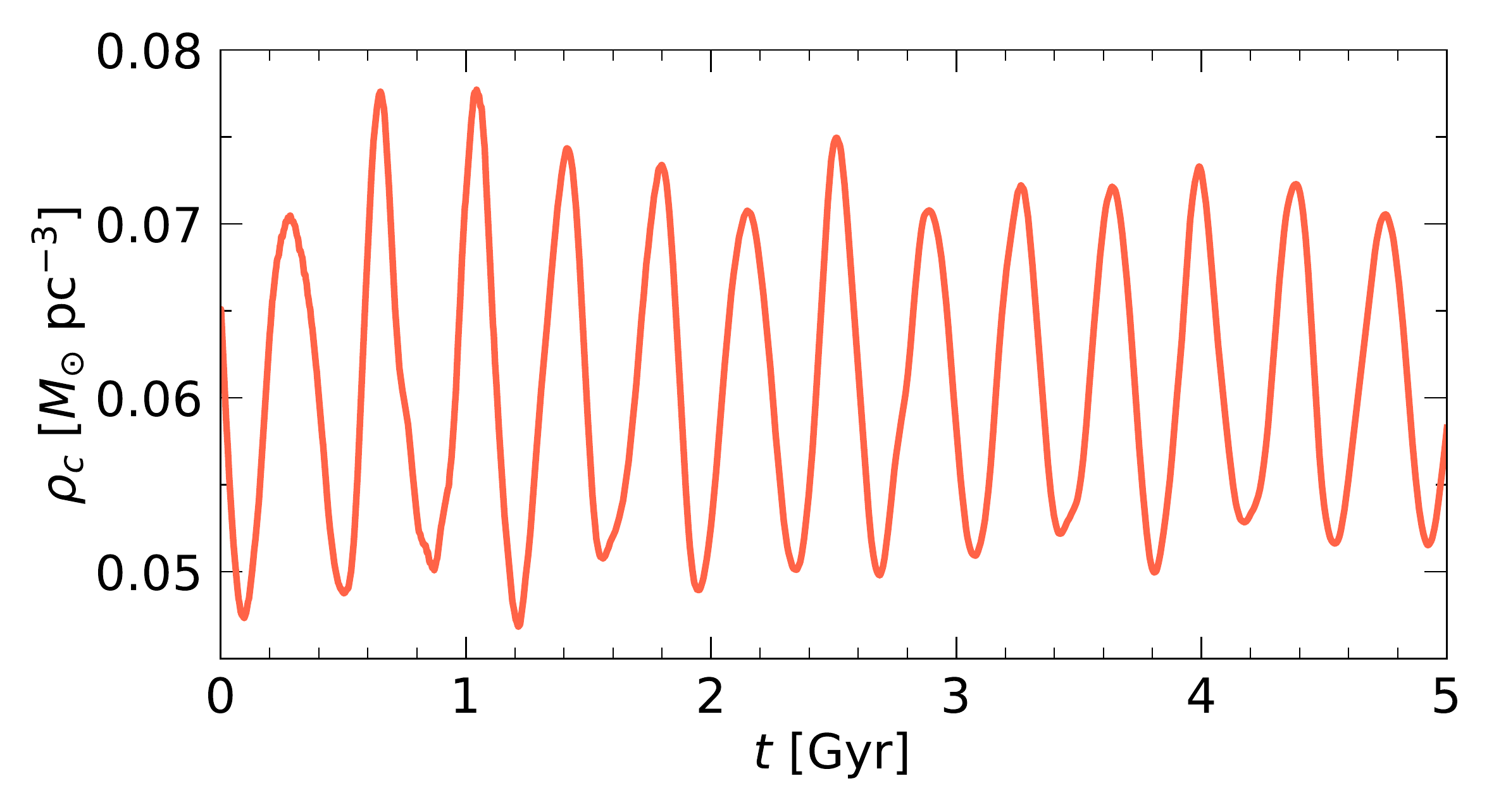}
	\caption{Time evolution of the peak soliton density in a self-consistent FDM simulation. The soliton oscillates with a characteristic period of $\tsol \sim 370\Myr$, consistent with \eref{eqn:NumFirstExcitedPeriod}. The oscillation amplitude is $C \sim 0.25$.}
	\label{fig:simu_dyn_sol-osc}
\end{figure}

Figure \ref{fig:simu_dyn_sol-osc} shows the time evolution of the peak soliton density. It demonstrates that the core oscillation is an intrinsic property that persists even in the absence of an outer halo. The gravitational influence of SC has no discernible impact on the density oscillations of the soliton, as $M_\text{soliton}/M_\text{SC} = \mathcal{O}(10^5)$. The observed oscillation amplitude is $C \sim 0.25$ and the oscillation period is $\tsol \sim 370\Myr$, in good agreement with \eref{eqn:NumFirstExcitedPeriod}.

\begin{figure}
	\includegraphics[width=\linewidth]{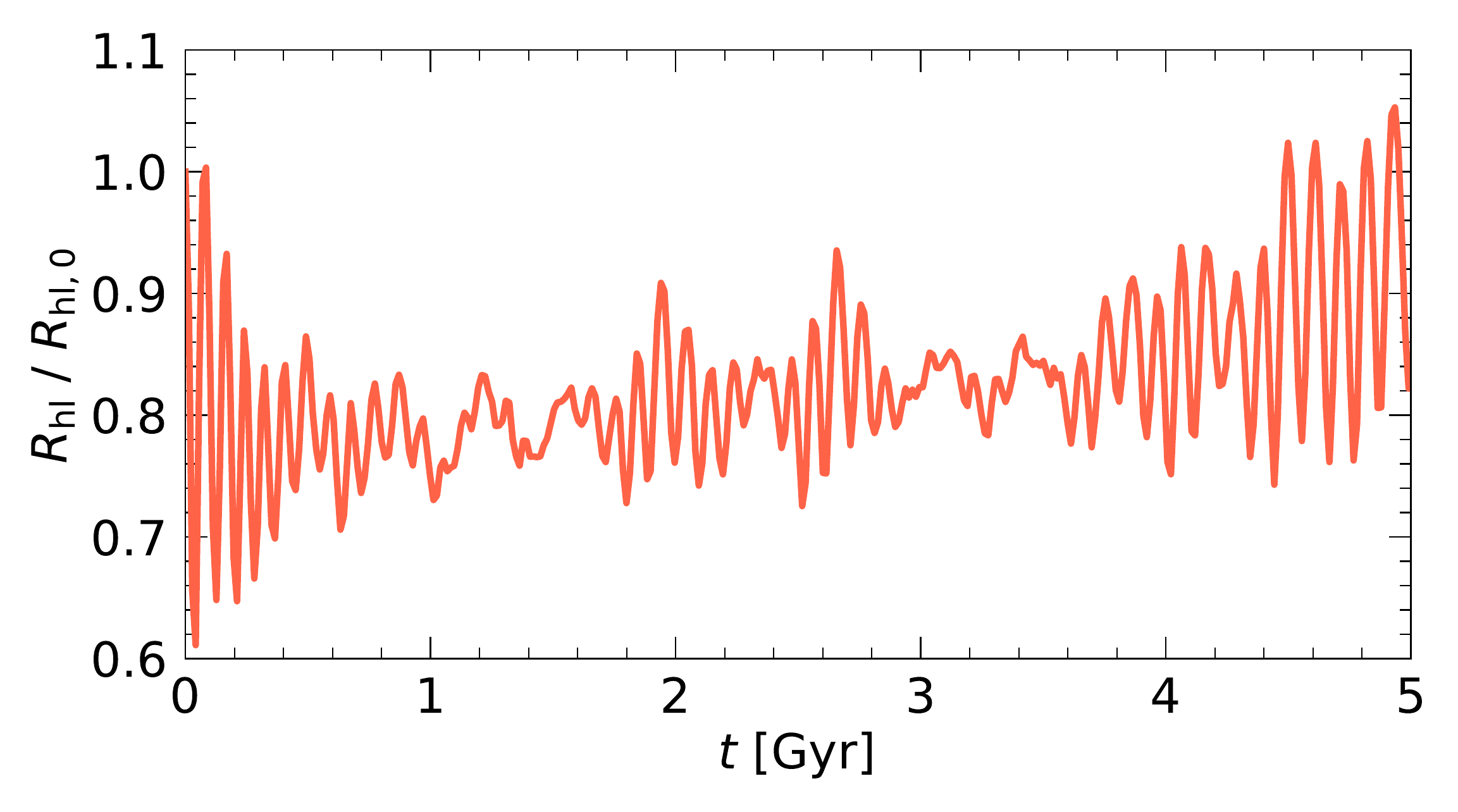}
	\caption{Time evolution of the SC's half-light radius $\Rhl$, normalized to its initial value $\Rhli$, in a self-consistent oscillating soliton. The effect of heating appears to be negligible.}
	\label{fig:simu_dyn_Rhl_evolution}
\end{figure}

\begin{figure}
	\includegraphics[width=\linewidth]{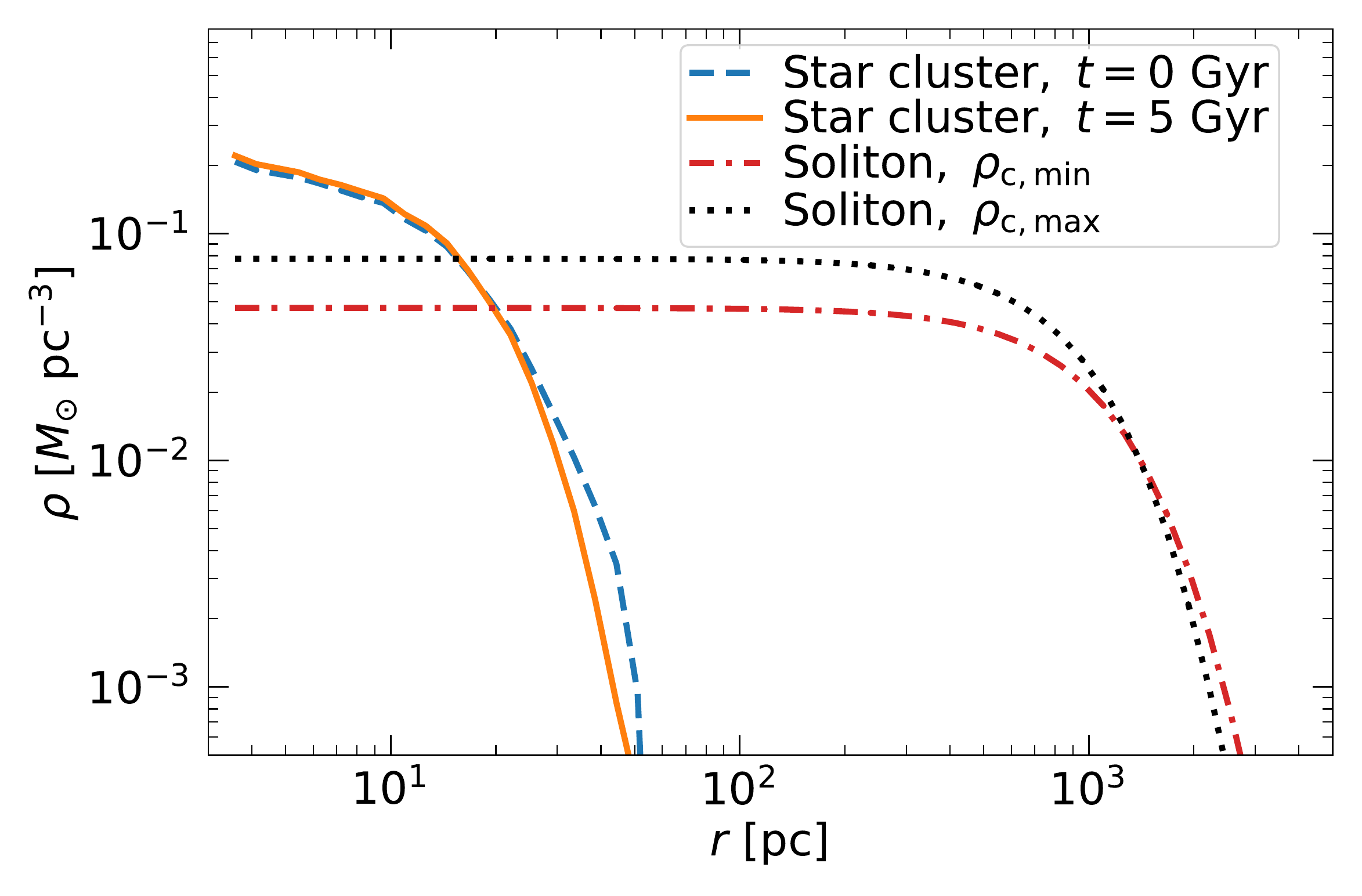}
	\caption{Density profiles in a self-consistent FDM simulation. The dashed and solid lines compare the stellar profiles at the beginning and end of the simulation, demonstrating that the SC is stable against an oscillating soliton. For reference, the dash-dotted and dotted lines show the soliton at its minimum and maximum densities.}
	\label{fig:simu_dyn_profile}
\end{figure}

Most importantly, we note that $\tsol/\tSC \sim 3$, suggesting negligible heating of the SC from the soliton oscillations (see also Figs. \ref{fig:NumericalRAverage} and \ref{fig:simu_ext_Rhl_max}). This is confirmed by \fref{fig:simu_dyn_Rhl_evolution}, which plots the projected half-light radius as a function of time. The maximum expansion of the half-light radius within $5\Gyr$ is found to be only $\Rhlmax/\Rhli \sim 1.05$. Figure \ref{fig:simu_dyn_profile} compares the stellar density profiles at $t=0$ and $5\Gyr$, further demonstrating that the SC is stable against the soliton oscillations.

\section{Comparison with PREVIOUS FDM Constraints from Eridanus II}\label{app:TragetPaper}
The constraints on FDM particle mass presented by \MN\, were set by the SC resonances, diffusion approximation, and subhalo mass function, together yielding $m_b \gtrsim 10^{-19}$ eV. In Sec. \ref{app:Comment1}, we discuss the dynamical timescales erroneously cited in \MN. We point out in Sec. \ref{app:Comment2} some mistakes and convergence issues with the perturbation calculations on gravitational heating presented therein. In Sec. \ref{app:Comment0}, we examine the validity of FDM constraints set by the subhalo mass function.

\subsection{Dynamical Timescales}\label{app:Comment1}

As stated in \MN, the diffusion approximation was applicable in the regime that satisfied $^\dagger\tau_\text{SC} \gg$ $^\dagger\tau_\text{soliton}$. In the limit $^\dagger\tau_\text{SC} \ll$ $^\dagger\tau_\text{soliton}$, they investigated SC heating, treating the SC as a virial system. The dynamical timescale of stochastic fluctuations (halo granules) was taken to be the same as $\tau_\text{soliton}$ in \MN. The validity of calculations therein rest first and foremost upon correctly determining the scaling relations of both $\tau_\text{SC}$ and $\tau_\text{soliton}$ as functions of $m_b$. The accurate analytical modeling of the soliton-SC system is equivalently crucial in yielding robust constraints.

The erroneous dynamical timescales of both the SC and an FDM soliton presented therein can be summarized as follows: (1) In computing $^\dagger\tau_\text{SC}$, the mass contribution of a soliton enclosed within $r_\text{SC}$ was completely neglected, which was only marginally correct for $m_b \lesssim 4.6\times 10^{-22}$ eV (albeit still with $\gtrsim 24\%$ error compared to $\tau_\text{SC}$, \eref{eqn:MassOrbitalPeriod}, in this limit). (2) The velocity dispersion of FDM $\sigma_\text{FDM}$ was assumed constant (i.e. independent of $m_b$) and equal to that of stars $\sigma_\text{EII}$, which was generally incorrect and led to the problematic scaling relation $^\dagger\tau_\text{soliton} \propto m_b^{-1}$. (3) The two distinct timescales of a soliton were conflated $^\dagger\tau_{00} =$ $^\dagger\tau_\text{soliton}$ in \MN. See Sec. \ref{sec:SolitonTimeScale} for discussions on this subtlety and Fig. \ref{fig:Gamer2PLOT2} for a comparison between the dynamical timescales adopted in \MN\, and this work; the discrepancy is further discussed below.

The dynamics of Eri II SC is determined gravitationally by the averaged mass enclosed within the orbit of individual member star. In \MN, the characteristic timescale of the SC was estimated by taking a test star on a Keplerian orbit with a semi-major axis $r_\text{SC}$: $^\dagger\tau_\text{SC} = 2\pi r_\text{SC}^{3/2}/(GM^{\leq r_\text{SC}}_\text{total})^{1/2}$, where the total enclosed mass $M^{\leq r_\text{SC}}_\text{total} = M_\text{soliton}^{\leq r_\text{SC}}+ M_\text{SC}^{\le r_\text{SC}}$ should include contributions from both the SC and soliton. For the fiducial $\rho_c = 0.15\Msun\pc^{-3}$ adopted in \MN, the enclosed soliton mass shall read $M_\text{soliton}^{\leq r_\text{SC}} \simeq 1380\Msun$ instead of the value $330\Msun$ quoted therein. The underestimated $M_\text{soliton}^{\leq r_\text{SC}}$ led to the incorrect approximation $M_\text{total}^{\leq r_\text{SC}} \sim M_\text{SC}^{\leq r_\text{SC}}$ (i.e. Eq. (18) in \MN). We have however demonstrated that $M_\text{soliton}^{\leq r_\text{SC}}/M_\text{SC}^{\le r_\text{SC}} \gtrsim \mathcal{O}(1)$ always holds regardless of the value of $m_b$ (see Fig. \ref{fig:SolitonMass}).

We have verified $\tau_\text{soliton} \simeq 2.3\tau_{00}$ in Sec. \ref{OscillationFrequencyRevisit}, as opposed to the assumption $^\dagger\tau_\text{soliton} = {^\dagger\tau_{00}}$ made by \MN. Operating under such an assumption, \MN\, then invoked the de Broglie relation to infer $^\dagger \tau_\text{soliton} = 2\pi \hbar/m_b \sigma_\text{FDM}^2$. The velocity dispersion of FDM was further assumed to be the same as that of Eri II, with $\sigma_{\text{1D,FDM}} = \sigma_\text{1D,EII} = 6.9 \text{ km s}^{-1}$ \cite{Li:2016utv}. The resulting scaling relation reads
\begin{eqnarray}
	^\dagger\tau_{00} &=& ^\dagger\tau_\text{soliton} \simeq 825\bigg(\frac{m_b}{10^{-22} \text{ eV}}\bigg)^{-1} \text{ Myr},
\end{eqnarray}
with a parameter dependence that is generally incorrect (cf. \eref{eqn:NumGroundPeriod}) \footnote{The parametric dependence of $\sigma_{\text{FDM}}$ can be understood heuristically by noting that $\tau_{00} \propto m_b^{-1}\sigma_{\text{FDM}}^{-2}$ and $\tau_{00} \propto \rho_c^{-1/2}$, yielding $\sigma_{\text{FDM}} \propto \rho_c^{1/4}m_b^{-1/2}$.}. For instance, when the physical size of a soliton gets larger than $r_\text{EII}$ for $m_b \lesssim 4.6 \times 10^{-22}$ eV, we have $\tau_\text{soliton} \sim$ const (i.e. independent of $m_b$; see \fref{fig:Gamer2PLOT2}). 

For $m_b > 10^{-20}$ eV, the condition $^\dagger\tau_\text{SC} \gg$ $^\dagger\tau_\text{granules}$ ensured the validity of diffusion approximation calculated in \MN, where the timescales of halo granular density fluctuations and of soliton oscillations were assumed identical $^\dagger\tau_\text{granules} \eee {^\dagger\tau}_\text{soliton}$. A comparable FDM constraint was derived in \cite{Amr2020} based on similar assumptions (e.g. anchoring the velocity dispersion of FDM to $\sigma_\text{EII}$). Our present analysis is not directly applicable to this large-mass regime, because the soliton oscillation timescale exhibits large uncertainty due to the unconstrained background halo profile modeling (gray shaded area in Fig.~\ref{fig:SolitonMass}). Additionally, the timescale of granular density fluctuations in relation to $\tau_\text{soliton}$ is also uncertain. Hence whether diffusion heating for $m_b > 10^{-20}\eV$ is as efficient as reported by \MN\, and \cite{Amr2020} remains to be investigated.

\subsection{Star Cluster Heating Calculations}\label{app:Comment2}
Gravitational heating caused by soliton density oscillations on the SC was investigated in \MN, where several (possibly) invalid assumptions were made in calculations therein. Our single-test-star model (\eref{eqn:TidalHeatingMass}) does not rely on these assumptions and yields more accurate predictions, consistent with both the N-body and FDM simulations.

To formulate the SC-soliton system, \MN\, first explicitly assumed the SC to be virialized for all time and have the unperturbed Hamiltonian $H_0 = G M_\text{SC}^{\le r_\text{SC}} m_\star/(2 a_0)$ for a star with mass $m_\star$ (Eq. (17) therein) with the semi-major axis $a_0 = r_\text{SC}$. The condition $^\dagger \tau_\text{SC} \ll$ $^\dagger\tau_\text{soliton}$ must be satisfied to be physically self-consistent, or otherwise we would need to demand $\mathcal{C} \ll 1$, given $M_\text{soliton}^{\leq r_\text{SC}}/M_\text{SC}^{\le r_\text{SC}} \gtrsim \mathcal{O}(1)$ holds true for all $m_b$ range of interest. Neither assumption is valid however, as shown in Sec. \ref{sec:SolitonTimeScale}. Here the mass contribution of the soliton was also erroneously neglected under the assumption $M_\text{soliton}^{\leq r_\text{SC}}/M_\text{SC}^{\le r_\text{SC}} \ll 1$. Furthermore, the soliton mass enclosed within $r_\text{SC}$ was also taken fixed (an incorrect assumption; see Sec. \ref{Heating}) to yield the perturbation Hamiltonian $\Delta H = \mathcal{C} [-G M_{\text{soliton}}^{\leq r_\text{SC}} m_\star/r(t)]\cos ({^\dagger\omega_\text{soliton}t})$, where $r(t)$ was the unperturbed orbit. The semi-major axis then evolved as
\begin{eqnarray}
	\dot{a} &=& 2\mathcal{C} \bigg(\frac{M_{\text{soliton}}^{\leq r_\text{SC}}}{M_\text{SC}^{\le r_\text{SC}}}\bigg) a_0^{2\dagger}\omega_\text{soliton}\frac{\sin ({^\dagger\omega_\text{soliton}t})}{r(t)}, \label{eqn:TidalResonanceIntegral}
\end{eqnarray}
which was integrated over $\text{Max}(^\dagger\tau_\text{orb}, {^\dagger\tau_\text{soliton}})$ with the initial condition $a = a_0$, fixing the eccentricity $e = 0.5$. The final semi-major axis $a_\text{final}$ was taken to be the average over $\tau_\text{ave} \eee10\times \text{Max}(^\dagger\tau_\text{orb}, {^\dagger\tau_\text{soliton}})$, and orbital resonances were considered efficient for $a_\text{final}/a_0 \geq 2$.

In this formulation, $^\dagger \tau_\text{SC}$ appeared only in the denominator of $\Delta H$ for elliptical orbits. For $e=0$, the solution of \eref{eqn:TidalResonanceIntegral} now depends solely on $^\dagger \tau_\text{soliton}$ as $r(t) = r_\text{SC} = \rm{const}$. Namely, $^\dagger\tau_\text{SC}$ becomes irrelevant in determining the efficiency of orbital resonances, which is unphysical.

Quantifying the potential perturbation induced by soliton density fluctuations by approximating $M^{\leq r(t)}_\text{soliton}$ as a point mass in $\Delta H$ effectively increases the oscillation amplitude $\mathcal{C}$ by a factor of two as detailed below. If one adopts the unperturbed soliton potential $V_0(\gamma; \rho_c, m_b)$, such an approach (perturbation Hamiltonian) to the first non-vanishing order in $\gamma$ yields (cf. \eref{eqn:appTidalHeating2ndODE})
\begin{eqnarray}\label{eqn:appMarsh}
	\ddot{r} =\omega_\text{SC}^2r- G\bigg\{\frac{M_\text{SC}}{r^2}+\frac{4\pi}{3}\rho_c r \Big[1+ 2\mathcal{C}\sin({^\dagger\omega_\text{osc}t+\phi)}\Big]\bigg\}.\nonumber\\
\end{eqnarray}
However, the density oscillation amplitude should be $\mathcal{C}$ instead of $2\mathcal{C}$ by construction (see \pAeref{eqn:SolitonOsc}\pBeref{eqn:TidalHeatingMass}). \Seref{eqn:appMarsh} in turn places an unphysical upper bound on $\mathcal{C} \leq 0.5$, above which $\rho_c$ is no longer strictly positive over a complete cycle of density oscillation.

Finally, in \MN, the resulting radial shifts, $a_\text{final}/a_0$, qualitatively diverge for other equivalently sensible choices of $\tau_\text{ave}$ (e.g. 3 Gyr). Similarly, varying the phase difference $\phi$ between the SC orbital motion and soliton density fluctuations, which was implicitly assumed to be zero in \MN, can also non-trivially alter the results for $e \neq 0$. In comparison, our approach does not suffer from these issues.

\subsection{Subhalo Mass Function}\label{app:Comment0}
In this subsection, we distinguish the unstripped halo mass at accretion $M_\text{halo}^{\leq r_\text{vir}}$ that obeys the core-halo mass scaling relation \eref{app:CoreHaloRelation} from the subhalo mass $M_\text{subhalo}^{\leq r_\text{vir}}$ after accounting for the effect of tidal stripping.

Given the subhalo mass function $d n_\text{sub}(m_b)/d \ln M_\text{subhalo}$, the predicted number of Milky Way subhalos $n_\text{EII}(m_b)$ reads
\begin{eqnarray}\label{eqn:SubhaloMassFunc}
	n_\text{EII}(m_b) = \int_{M_\text{subhalo}^{-2\sigma}}^{M_\text{subhalo}^{+2\sigma}} d \ln M_\text{subhalo} \bigg[\frac{dn_\text{sub}(m_b)}{d \ln M_\text{subhalo}}\bigg],\;\;\;\;\;\;
\end{eqnarray}
within 95\% confidence level (CL) of the Eri II halo mass $M_\text{subhalo}^{\leq r_\text{vir}}$, where $r_\text{vir}$ denotes the halo virial radius. The existence of Eri II asserts that $n_\text{EII}(m_b) \geq 1$; by directly equating $M_\text{subhalo}^{\leq r_\text{vir}}$ to $M_\text{EII}^{\leq r_\text{EII}} = 1.2^{+0.4}_{-0.3} \times10^7$ M$_\odot$ at 68\% CL, a lower bound $m_b \gtrsim 8\times 10^{-22}$ eV was derived by \MN.
\begin{figure}
	\includegraphics[width=\linewidth]{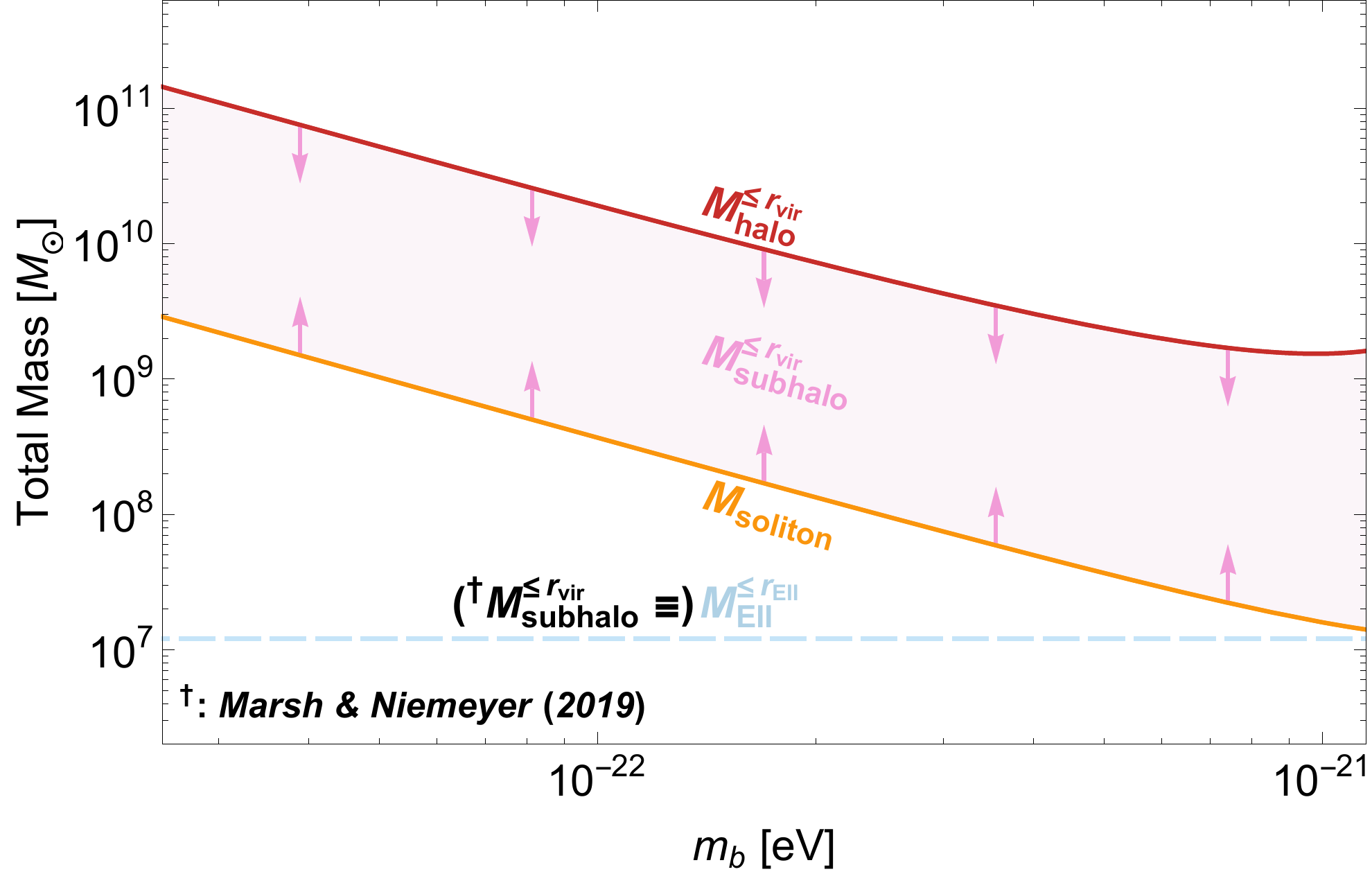}
	\caption{Unstripped halo mass $M_\text{halo}^{\leq r_\text{vir}}$ (red) and soliton mass (orange) consistent with both observations (Fig. \ref{fig:SolitonMass}) and \eref{eqn:solitonhalomassratio}. The allowed values of subhalo mass $M_\text{subhalo}^{\leq r_\text{vir}}$ (pink shaded) are compared with that adopted by \MN\, \cite{Marsh:2018zyw}, which was set to be the enclosed mass of Eri II within its half-light radius $^\dagger M_{\rm subhalo}^{\le r_{\rm vir}} \eee M_\text{EII}^{\leq r_\text{EII}}$ (light blue). Note that in general $M_{\rm subhalo}^{\le r_{\rm vir}} \gg M_\text{EII}^{\leq r_\text{EII}}$.}
	\label{fig:MhaloFunction}
\end{figure}

One major issue of the subhalo mass function calculation by \MN\, lies in the questionable assumption of $M_\text{subhalo}^{\leq r_\text{vir}} = M_\text{EII}^{\leq r_\text{EII}}$. From the core-halo relation and the observed mean density of Eri II within $r_\text{EII}$, we have $M_\text{halo}^{\leq r_\text{vir}} \gg M_\text{EII}^{\leq r_\text{EII}}$ as detailed below. First note the soliton mass contribution always dominates within $3.3 r_c$ compared with that of a background halo, as noted in Sec. \ref{sec:AstroConstraints}. The core-halo mass relation \cite{Schive:2014hza} and the total soliton mass (\eref{eqn:SolitonTotalMass}) together yield the soliton-halo mass ratio at redshift zero (see Appendix \ref{app:SolitonUnperturbed})
\begin{eqnarray}	
	\frac{M_\text{halo}^{\leq r_\text{vir}}}{M_\text{soliton}} = 135\bigg(\frac{\rho_c}{\text{M}_\odot \text{pc}^{-3}}\bigg)^{1/2}, \label{eqn:solitonhalomassratio}
\end{eqnarray}
which carries no (explicit) dependence on $m_b$. $\rho_c \geq 0.135$ M$\odot$ pc$^{-3}$ for Eri II then implies $M_\text{halo}^{\leq r_\text{vir}} \geq 49.6M_\text{soliton}$. As demonstrated in Fig. \ref{fig:SolitonMass}, we have $3.3 r_c \geq r_{\rm EII}$ for $m_b \lesssim 10^{-21}$ eV, and thus $M_\text{halo}^{\leq r_\text{vir}} \gg M_\text{soliton}^{\leq 3.3 r_c} \geq M_\text{EII}^{\leq r_\text{EII}}$. Figure~\ref{fig:MhaloFunction} compares the values of $M_\text{halo}^{\leq r_\text{vir}}$ (red), $M_\text{soliton}$ (orange), and $M_\text{EII}^{\leq r_\text{EII}}$ that are consistent with both observations and \eref{eqn:solitonhalomassratio}. The maximum $m_b$ plotted, $\sim 10^{-21}\eV$, covers the lower bound $m_b \gtrsim 8\times10^{-22}\eV$ derived by \MN. Evidently, $M_\text{halo}^{\leq r_\text{vir}} \gg M_\text{soliton} \gtrsim M_\text{EII}^{\leq r_\text{EII}}$ for the entire range of $m_b$ of interest. Moreover, in our cosmological simulations, Eri-II-like halos (i.e. with $\rho_c \sim 0.135$ M$_\odot$pc$^{-3}$) do naturally form with $m_b \sim 10^{-22}$ eV \cite{Schive:2014dra, Schive:2014hza}, which further supports the consistency of the preceding analysis.

An accurate estimate of subhalo mass $M_\text{subhalo}^{\leq r_\text{vir}}$ requires detailed information on the soliton-subhalo evolution history in response to the tidal field of the host galaxy. The pink shaded region in Fig. \ref{fig:MhaloFunction} indicates observationally compatible values of $M_\text{subhalo}^{\leq r_\text{vir}}$. In more extreme cases with substantial tidal stripping, the subhalo mass $M_\text{subhalo}^{\leq r_\text{vir}}$ can reduce to a mass scale comparable to $M_\text{soliton}$ \cite{Du:2018qor, Schive:2019rrw}. Here we have assumed that the soliton remains intact since a partially stripped soliton is unstable \cite{Du:2018qor}. Therefore, the inequality $M_\text{halo}^{\leq r_\text{vir}} \geq M_\text{subhalo}^{\leq r_\text{vir}} \geq M_\text{soliton} \gg M_\text{EII}^{\leq r_\text{EII}}$ always holds for $m_b < 8\times10^{-22}$ eV, even when assuming complete tidal disruption of the surrounding halo. The revised subhalo mass thus frees up the previously claimed lower bound of particle mass $m_b \gtrsim 8\times10^{-22}\eV$ by \MN.

The other questionable aspect of \eref{eqn:SubhaloMassFunc} lies in the integration over the uncertainty range $\pm 2\sigma$, which gives $n_\text{EII} \rightarrow 0$ when $\sigma \rightarrow 0$ irrespective of the subhalo mass function. A more sensible approach is to either include more halo samples to estimate the halo number density around $M_\text{subhalo}^{\leq r_\text{vir}}$ or calculate the expectation value of $n_\text{EII}(<M_\text{subhalo}^{+2\sigma})$ from the cumulative subhalo mass function. The halo-to-halo variance should also be properly accounted for.

\section{Summary and Conclusions}\label{sec:conclude}
We have presented a detailed analytical and numerical study of soliton oscillations and identified two characteristic timescales associated, respectively, with the ground-state soliton wavefunction $\tau_{00}$ and the soliton density oscillations $\tau_\text{soliton}$; the astrophysical constraints from Eri II have also been reexamined. The main results are enumerated as follows.
\begin{itemize}
	\item Both $\tau_{00}$ and $\tau_\text{soliton}$ depend only on the soliton peak density $\rho_c$, with a constant ratio $\tau_\text{soliton}/\tau_{00} \simeq 2.3$ irrespective of the boson mass $m_b$.
	\item The timescale of the central star cluster (SC) in Eri II, $\tau_\text{SC}$, has $\tau_\text{soliton} / \tau_\text{SC} \simeq 2\textendash3$ that deviates substantially from unity.
	\item Star cluster heating by an oscillating external gravitational field with a period $\tau_\text{osc}$ is noticeable only for $\tau_\text{osc}/\tau_\text{SC} \simeq 0.5$\textendash$1$. The heating by an oscillating soliton is hence ineffective, given $\tau_\text{soliton} / \tau_\text{SC} \simeq 2\textendash3$.
	\item Some of the FDM constraints derived by \MN\, appear invalid. The effect of orbital resonances is negligible over 10 Gyr as detailed above. The constraint from subhalo mass function cited therein is overly stringent, derived from a severely underestimated Eri II subhalo mass.
\end{itemize}

We have implicitly assumed in this work
that the background halo of Eri II has been stripped away by the tidal field
of the Milky Way. Otherwise, the presence of a background halo would lead to
soliton random walk excursions, causing complete tidal disruption of the SC
within $\sim$ 1 Gyr for $m_b \sim 10^{-22}\eV$ \cite{Schive:2019rrw}.
Accordingly, the analysis here is mainly applicable for
$m_b \lesssim 3.8\times10^{-21}\eV$, above which the halo contribution becomes
non-negligible and requires further investigation.

We have only used the average mass density within the half-light radius of
Eri II ($\rho_\text{EII}$) to constrain the soliton profile, which however
cannot break the degeneracy between $\rho_c$ and $m_b$.
This limitation can be relaxed using the dynamics of Eri II member stars to
determine the enclosed mass profile (e.g. see \cite{Zoutendijk:2021kee}).
In addition, further observations capable of accurately determining the
velocity dispersion of Eri II SC will conclude whether the SC is self-bound
(or not), indicating $m_b \lesssim (\gtrsim) 4.6 \times 10^{-22}\eV$
(see Fig. \ref{fig:SolitonMass}). Lastly, we note that the tidal disruption of
the SC due to soliton random walk excursions should be investigated more
extensively with different $m_b$ and $M_{\rm halo}$.

The particle mass constraint from the subhalo mass function discussed in
\sref{app:Comment0} focuses on a specific halo mass range corresponding to
Eri II. Nadler et al. \cite{Nadler:2020prv}, on the other hand, used the
Milky-Way satellite galaxy luminosity functions, especially the faintest
galaxies, to obtain a stringent constraint of $m_b \gtrsim 2.9 \times 10^{-21}\eV$.
Cosmological FDM simulations capable of forming Milky Way-sized halos
are indispensable for investigating this issue further, especially regarding
whether the smaller Jeans wavelength and the extra small-scale powers
developed at lower redshifts (e.g. see \cite{2021arXiv210101828M})
could increase the abundance of low-mass halos.

\subsection*{Acknowledgments}
We are happy to acknowledge useful conversations with Frank van den Bosch, Dhruba Dutta Chowdhury, Zhi Li, David J. E. Marsh, Jens C. Niemeyer, and Daniel W. Boutros. H. S. acknowledges funding support from the Jade Mountain Young Scholar Award No. NTU-109V0201, sponsored by the Ministry of Education, Taiwan. This work is partially supported by the Ministry of Science and Technology (MOST) of Taiwan under Grants No. MOST 107-2119-M-002-036-MY3 and No. MOST 108-2112-M-002-023-MY3, and the NTU Core Consortium project under Grants No. NTUCC-108L893401 and No. NTU-CC-108L893402.

\appendix

\section{Attributes of an Unperturbed Soliton}\label{app:SolitonUnperturbed}
\subsection{Ground-state Soliton Potential}\label{app:GravitationalPotential}
The self-interaction gravitational potential of the coupled SP equations (\pCAeref{eqn:SPeqnS}\pCCeref{eqn:SPeqnP}) is specified by the wavefunction $\Psi$ of interest. The spherical symmetry of the ground-state soliton potential grants the separation of wavefunctions into radial eigenfunctions and spherical harmonics $Y_l^m(\theta, \psi)$. We adopt the following decomposition in spherical coordinates ($n \in \mathbb{Z}^{0+}$)
\begin{eqnarray}\label{eqn:WaveFDecomposition}
	\begin{cases}
		\Psi(\gamma,\theta, \phi, t) = \sum c_{nlm} \psi_{nlm}(\gamma, \theta, \psi) e^{-i E_{nl} t/\hbar},\\
		\psi_{nlm}(\gamma, \theta, \phi) = (r_c \gamma)^{-1}u_{nl}(\gamma) \times Y_l^m(\theta, \phi),
	\end{cases}
\end{eqnarray}
where $r_c \int_{0}^{\infty} |u_{nl}(\gamma)|^2 d\gamma = 1$. The ground-state wavefunction $\psi_{000}$ is an exact solution to the unperturbed SP equations; the Poisson equation reads
\begin{eqnarray}
\nabla^2 V_0 = 4\pi G m_b |\psi_{000}|^2= 4\pi G \rho_\text{soliton}(\gamma; \rho_c),
\end{eqnarray}
where the density distribution is given in \eref{eqn:gammaDensityDist}.
The del operators in spherical coordinates acting on scalar function $f(\gamma)$ with only radial dependence are
\begin{eqnarray}
\begin{cases}\label{eqn:DelOperators}
\nabla f= \frac{\partial f}{\partial r} \hat{\mathbf{r}}= r_c^{-1} \frac{\partial f}{\partial \gamma} \hat{\mathbf{r}},\\
\nabla^2f = r_c^{-2} \frac{1}{\gamma}\frac{\partial^2}{\partial \gamma^2}(\gamma f )=r_c^{-2} \frac{1}{\gamma^2}\frac{\partial}{\partial \gamma}\Big(\gamma^2\frac{\partial f}{\partial \gamma}\Big) .
\end{cases}
\end{eqnarray}
The gravitational potential (per unit mass) of an unperturbed soliton can then be solved analytically:
\begin{eqnarray}
 V_0 &=& -\frac{625\pi G r_c^2\rho_c}{794976}\Bigg\{99\sqrt{910}\gamma^{-1}\times \arctan{\bigg(\frac{1}{10}\sqrt{\frac{91}{10}}\gamma\bigg)}\nonumber \\
&&+\bigg[\frac{1.69\times10^4}{(10^3+91\gamma^2)^6}\bigg](1.83\times 10^{17}+5.136\times 10^{16}\gamma^2 \nonumber \\
&& +7.088\times 10^{15}\gamma^4+5.303\times10^{14}\gamma^6\nonumber \\
&&+2.072\times 10^{12}\gamma^{8}+3.327\times 10^{11} \gamma^{10}) \Bigg\}, \label{eqn:AppVExcitedStatesAnalysis}
\end{eqnarray}
which is required to vanish at infinity.
\begin{figure}
	\includegraphics[width=\linewidth]{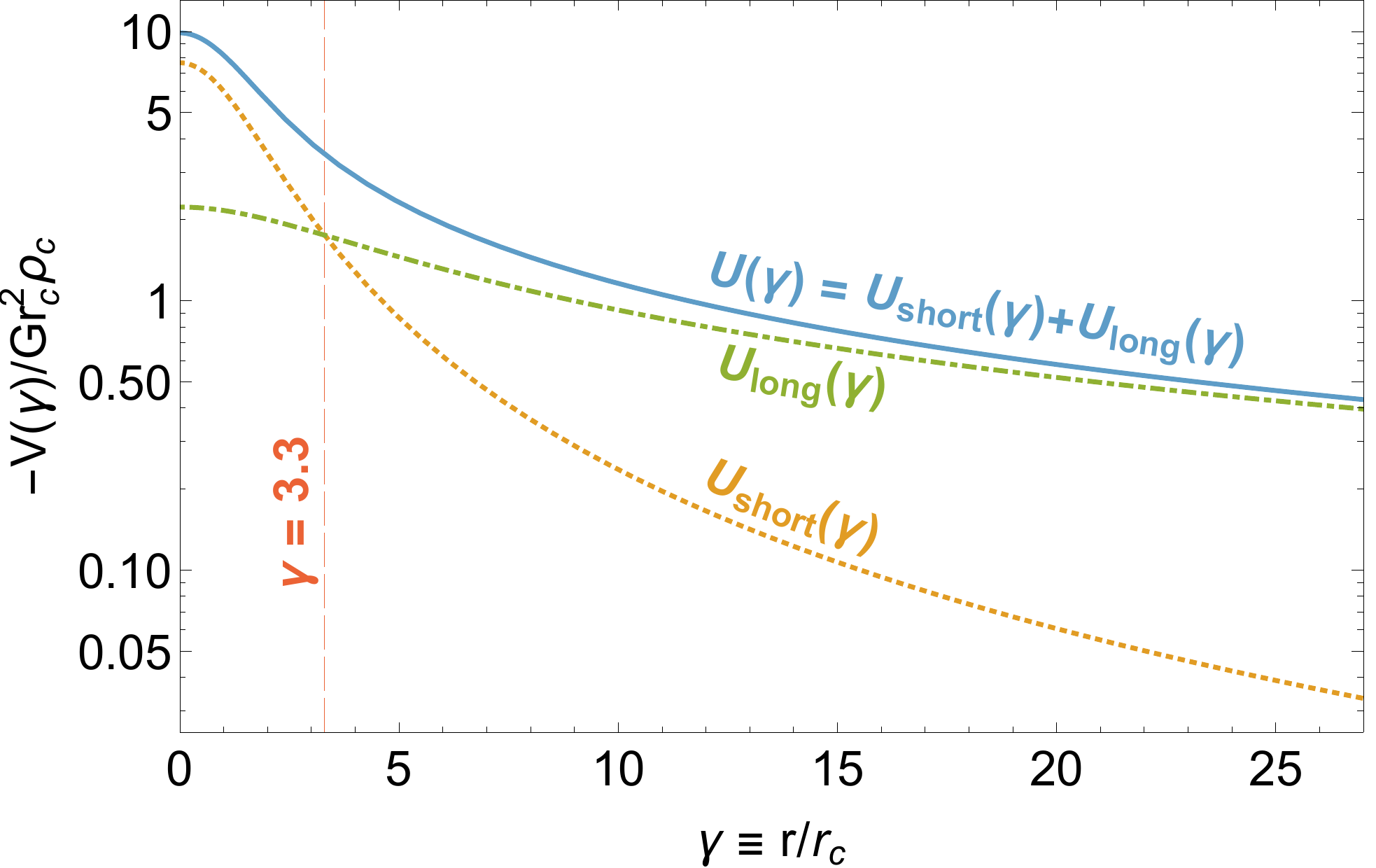}
	\caption{Scaled gravitational potential of a unperturbed FDM soliton. The transition between the long-range and short-range decompositions (\pAeref{eqn:VGeneric}\pBeref{eqnSolitonPotentialDecomp}) occurs at $\gamma = 3.3$.}
	\label{fig:SolitonPotential}
\end{figure}

The parametric dependence of the ground-state soliton potential can be separated
\begin{eqnarray}
V_0(\gamma; \rho_c, m_b)&=& - G r_c^2 \rho_c U(\gamma),\;\;\;\;\;\;\;\;
\label{eqn:VGeneric}
\end{eqnarray}
and decomposed into a short-range term $U_\text{short}(\gamma)$ (that dominates for $\gamma<3.3$) and a long-range piece $U_\text{long}(\gamma)$ (for $\gamma>3.3$), where
\begin{eqnarray}
\begin{cases}
U(\gamma) = U_\text{short}(\gamma)+U_\text{long}(\gamma)\\
U_\text{short}(\gamma) \eee
\Big[\frac{4.17\times 10^{-17}}{(1+9.1\times 10^{-2}\gamma^2)^6}\Big](1.83\times 10^{17} \\+5.14\times 10^{16}\gamma^2+7.09\times 10^{15}\gamma^4 \\+ 5.30\times 10^{14}\gamma^6+2.07\times 10^{12}\gamma^{8}+3.33\times 10^{11} \gamma^{10}),\\
U_\text{long}(\gamma)\eee 7.38\gamma^{-1} \arctan(0.302\gamma),\label{eqnSolitonPotentialDecomp}
\end{cases}
\end{eqnarray}
and $U_\text{short}(\gamma)/U_\text{long}(\gamma)|_{\gamma=3.3}=1$. Figure \ref{fig:SolitonPotential} plots $U_\text{short}$ (dotted yellow) and $U_\text{long}$ (dash-dotted green) at radius $\gamma$ from the center of a soliton. The soliton core behaves quantitatively as a point mass for $r > 3.3 r_c$; the potential approaches zero as $\gamma \rightarrow \infty$. In the limit $\gamma \ll 1$, \eref{eqn:AppVExcitedStatesAnalysis} has the following asymptotic expression
\begin{eqnarray}
V_0(\gamma; \rho_c, m_b) &\simeq& -G r_c^2 \rho_c\big(9.86-2.09\gamma^2\big), \label{eqn:PotentialSmallgApprox}
\end{eqnarray}
accurate to $0.01 \%$ for $\gamma \leq 0.1$ ($4 \%$ for $\gamma \leq 1$).
The soliton potential at the center, relative to $V_0(\gamma \rightarrow \infty) = 0$, is
\begin{eqnarray}
m_b V_0(0;\rho_c, m_b) &=& -6.51 \times 10^{-30} \bigg(\frac{\rho_c}{\text{M}_\odot \text{pc}^{-3}}\bigg)^{1/2} \text{eV},\;\;\;\;\;\;\:\:\label{eqn:SolitonPotentialCenter}
\end{eqnarray}
which depends only on $\rho_c$.

\subsection{Soliton Energy Levels}\label{app:SolitonEnergyLevels}
The radial Schr\"{o}dinger equation takes the form
\begin{eqnarray}\label{eqn:RadialSEq}
\frac{\partial^2 u_{nl}(\gamma)}{\partial \gamma^2} &=& \Bigg[\frac{2 r_c^2 m_b}{\hbar^2}\big(m_b V_0-E_{nl}\big) +\frac{l(l+1)}{\gamma^2}\Bigg]u_{nl}(\gamma),\nonumber\\\;
\end{eqnarray}
where $r_c^2 m_b \propto \rho_c^{-1/2}$. Note that the total energy, radial eigenfunction, and associated oscillation frequency for different (excited) states depend only on the FDM central density $\rho_c$. We impose a homogeneous boundary condition and numerically solve for the eigenfunctions and associated eigenvalues of the system.

Table \ref{tab:SolitonTotalEnergy} lists the total energy $\frac{E_{nl}(\rho_c)}{|m_b V_0(0; \rho_c, m_b)|}$ normalized to the soliton potential at the center (\eref{eqn:SolitonPotentialCenter}) for each soliton energy state $(n,l)$. The ground-state soliton is well virialized, consistent with zoom-in simulations \cite{Veltmaat:2018dfz}. The oscillation periods of the ground-state soliton wavefunction obtained analytically
\begin{eqnarray}
\tau_{00}(\rho_c)&=& \frac{2\pi \hbar}{|E_{00}|} \simeq 39.9 \bigg(\frac{\rho_c}{\text{M}_\odot \text{pc}^{-3}}\bigg)^{-1/2} \text{ Myr}\;\;\;\;\;
\end{eqnarray}
and observed in simulations \cite{Schive:2019rrw} differ less than $5\%$.
\begin{table}
	\begin{center}
		\begin{tabular}{|c|c|c|c|c|}
			\hline$\frac{E_{nl}(\rho_c)}{|m_b V_0(0; \rho_c, m_b)|}$
			&$ l=0 $ & $ l=1 $ & $ l=2 $ & $l=3$ \\\hline 	\hline
			$n=0$ & $-0.505$ &$-$ &$-$ & $-$\\	\hline
			$n=1$ & $-0.198$ &$-0.125$ &$-$ & $-$ \\\hline
			$n=2$ & $-0.103$ &$-0.0756$ &$-0.0543$ &$-$ \\\hline
			$n=3$ & $-0.0654$ &$-0.0509$ &$-0.0386$ & $-0.0289$\\\hline
			$n=4$ & $-0.0452$ &$-0.0365$ &$-0.0288$ & $-0.0223$\\\hline
		\end{tabular}
	\end{center}
	\vspace*{-1mm}
	\caption{Scaled total energy $\frac{E_{nl}(\rho_c)}{|m_b V_0(0; \rho_c, m_b)|}$ of a soliton eigenstate $(n,l)$. The ground state soliton appears to be well virialized, which was previously observed in soliton-halo simulations \cite{Veltmaat:2018dfz}.}
	\label{tab:SolitonTotalEnergy}
\end{table}

\subsection{Soliton Profile and Soliton-Halo Mass Ratio}
The mass of an unperturbed soliton enclosed within the scaled radius $\gamma$ has \footnote{This expression is equivalent to Eq. (A1) in \cite{Chen:2016unw}. The $3\%$ difference in value results from rounding errors in \pAeref{eqn:gammaDensityDist}\pBeref{eqn:SolitonRadiusEriII} with coefficients $1.9$ ($1.95$) and $9.1\times 10^{-2}$ ($9.06 \times 10^{-2}$).}
\begin{eqnarray}\label{eqn:SolitonEncMass}
	&&M_\text{soliton}^{\leq \gamma}(\gamma; \rho_c, m_b) = 4\pi r_c^3 \int_{0}^{\gamma}d\gamma' \big[\gamma'^2 \rho_\text{soliton}(\gamma';\rho_c)\big]\\ &&= 7.376\rho_c r_c^3\times \Big[\arctan\big(0.3017 \gamma\big)+\big(1+9.1\times 10^{-2}\gamma^2\big)^{-7}\nonumber\\&&\times(-0.3017 \gamma + 0.3849 \gamma^3 + 0.06656 \gamma^5 + 6.651\times 10^{-3} \gamma^7 \nonumber\\&&+
	3.903\times 10^{-4} \gamma^9 + 1.255\times 10^{-5} \gamma^{11} + 1.713\times 10^{-7} \gamma^{13})\Big]. \nonumber
\end{eqnarray}
The total mass reads (taking $\gamma \rightarrow \infty$)
\begin{eqnarray}
	M_\text{soliton}(\rho_c, m_b) &=& 11.6 \rho_c r_c^3. \label{eqn:SolitonTotalMass}
\end{eqnarray}
To first non-vanishing order in $\gamma$ (valid for $\gamma \ll 1$), we recover the expression of total enclosed mass in a background of uniform mass distribution ($4\pi/3 = 4.19$)
\begin{eqnarray}
	M_\text{soliton}^{\leq \gamma}(\gamma; \rho_c, m_b) = 4.19\rho_c r_c^3\gamma^3. \label{eqn:SolitonEncMassSmallRadius}
\end{eqnarray}
The normalized radial eigenfunction $u_{nl}$ for a spherically symmetric configuration (i.e. $(n,l,m) = (n,0,0), \forall n, l \in \mathbb{Z}^{0+}$) and the corresponding density distribution $\rho_{n0}$ are related via the Poisson equation:
\begin{eqnarray}	
	r_c \int_0^\gamma d\gamma' |u_{n0}(\gamma')|^2 = \frac{4\pi r_c^3 \int_0^\gamma d\gamma' \Big[\gamma'^2 \rho_{n0}(\gamma')\Big]}{4\pi r_c^3 \int_0^\infty d\gamma' \Big[\gamma'^2 \rho_{n0}(\gamma')\Big]}, \;\;\;\;\;\;\;\label{eqn:WavefunDen}
\end{eqnarray}	
where we notice that the soliton total mass
\begin{eqnarray}	
	M_{\text{soliton},n0} \eee 4\pi r_c^3 \int_0^\infty d\gamma' \Big[\gamma'^2 \rho_{n0}(\gamma')\Big],
\end{eqnarray}	
is constant for fixed $\rho_c$ and $r_c$, irrespective of energy state $(n,l)$; namely, $M_{\text{soliton},nl}=M_\text{soliton}(\rho_c, r_c), \forall n,l \in \mathbb{Z}^{0+}$. This gives
\begin{eqnarray}	
	|u_{n0}(\gamma)|^2 = \frac{4\pi r_c^2 \Big[\gamma^2 \rho_{n0}(\gamma)\Big]}{M_{\text{soliton}}}. \label{eqn:appDensityWF}
\end{eqnarray}	

The total mass of an FDM halo within its virial radius $r_\text{vir}$ at redshift zero is uniquely specified by $M_\text{soliton}$ via the core-halo relation \cite{Schive:2014hza}
\begin{eqnarray}\label{app:CoreHaloRelation}
	M_\text{soliton} = \frac{M_0}{4}\bigg(\frac{M_\text{halo}}{M_0}\bigg)^{1/3},
\end{eqnarray}	
where $M_0 \simeq 4.4\times 10^7(m_b/10^{-22} \text{eV})^{-3/2}$ M$_\odot$. The core mass (\eref{eqn:SolitonTotalMass}) together with \eref{eqn:SolitonRadiusEriII} can then be plugged in to yield
\begin{eqnarray}	
	M_\text{halo} = \bigg(\frac{\rho_c}{1.9 \text{ M}_\odot \text{pc}^{-3}}\bigg)^{3/2}\bigg(\frac{r_c}{62.5 \text{ pc}}\bigg)^3 10^9 \text{ M}_\odot.\;\;\;\;\;\;\;\;\label{eqn:HaloSolitonRatio}
\end{eqnarray}
The soliton-halo mass ratio reads
\begin{eqnarray}	
	\frac{M_\text{halo}(\rho_c, m_b)}{M_\text{soliton}(\rho_c, m_b)} = 135\bigg(\frac{\rho_c}{\text{M}_\odot \text{pc}^{-3}}\bigg)^{1/2}, \label{app:solitonhaloRatio}
\end{eqnarray}
which depends only on $\rho_c$. Note that here we make no assumption on the form of the FDM halo profile.

An FDM halo is well approximated by a Navarro-Frenk-White (NFW) profile \cite{Schive:2014dra, Schive:2014hza, Mocz:2017wlg, Pozo:2020ukk}
\begin{eqnarray}
	\rho_h(\gamma;\rho_0,r_s) = \frac{\delta_h \rho_0}{\gamma \times\frac{r_c}{r_s}\Big(1+\gamma \times\frac{r_c}{r_s}\Big)^2}, \label{eqn:HaloProfile}
\end{eqnarray}
where the halo density contrast $\delta_h$ and the scale radius $r_s$ are conventionally treated as free parameters. Adopting $H_0 = 67.7$ km s$^{-1}$ Mpc$^{-1}$ \cite{Aghanim:2018eyx}, the critical density at redshift zero has
\begin{eqnarray}
	\rho_0 \eee \bigg(\frac{3H_0^2}{8\pi G}\bigg) = 1.27\times 10^{-7} \text{ M}_\odot\text{pc}^{-3}.
\end{eqnarray}
The density contrast is defined as \cite{Navarro:1995iw, Navarro:1996gj}
\begin{eqnarray}
	\delta_h &\eee& \Big(\frac{200}{3}\Big) \Bigg[\frac{c^3}{ \ln(1+c)-\frac{c}{1+c}}\Bigg], \label{eqn:DensityConstrast}
\end{eqnarray}
where the value of concentration parameter $c = 4$\textemdash$40$ depends on the formation history of the galaxy under study. The halo mass enclosed within the virial radius reads
\begin{eqnarray}
	M_{h} =4\pi \delta_h \rho_0 r_s^3 \bigg[\ln(1+c)-\frac{c}{1+c}\bigg]. \label{eqn:NFWHaloMass}
\end{eqnarray}
Combing \pCAeref{eqn:SolitonTotalMass}\pCBeref{eqn:HaloSolitonRatio}\pCCeref{eqn:NFWHaloMass}, we obtain the concentration parameter
\begin{eqnarray}
	c^3  =1.87 \bigg(\frac{r_c}{r_s}\bigg)^3 \bigg(\frac{\rho_c}{\rho_0}\bigg)\bigg(\frac{\rho_c}{\text{ M}_\odot \text{pc}^{-3}}\bigg)^{1/2}, \label{eqn:crhrcRelation}
\end{eqnarray}
which at redshift zero reduces to
\begin{eqnarray}
	c = 245\bigg(\frac{\rho_c}{\text{ M}_\odot\text{pc}^{-3}}\bigg)^{1/2}\bigg(\frac{r_c}{r_s}\bigg). \label{eqn:rhohGeneric}
\end{eqnarray}
The density contrast $\delta_h(\rho_c, \eta)$ is thus completely specified by $r_s$ for a given soliton (fixed $\rho_c$ and $r_c$). The FDM halo profile effectively reduces to an one-parameter scaling relation for a given soliton. Here we emphasize that the soliton-halo mass ratio \eref{app:solitonhaloRatio} is independent of halo scaling parameters.

\section{Time-dependent Solitary Perturbations}\label{app:PerturbationAppendix}

\subsection{First-order Density Perturbation}\label{app:PerturbationI}
The spherical symmetry of the ground-state soliton wavefunction motivates us to consider density perturbations isotropic about the center of mass of a soliton:
\begin{eqnarray}
\bigg(-i\hbar \frac{\partial}{\partial t} -\frac{\hbar^2}{2m_b} \nabla^2 + m_bV_0\bigg)\frac{\delta u(\gamma, t)}{r_c \gamma} Y_0^0(\theta, \phi) =\;\;\;\;&&\nonumber\\ -m_b\delta V \frac{u_{00}(\gamma)}{r_c \gamma}Y_0^0(\theta, \phi) e^{-i E_{00}t/\hbar}&&,\;\;\;\;\;\;\;\;\;\; \label{eqn:AppRadialPerturbationEqUns}
\end{eqnarray}
where $Y_0^0(\theta,\phi)=(\sqrt{1/\pi})/2$ is a constant. We can easily expand the Laplacian on the LHS of \eref{eqn:AppRadialPerturbationEqUns}
\begin{eqnarray}
\nabla^2\bigg(\frac{\delta u(\gamma, t)}{r_c \gamma} Y_0^0(\theta, \phi)\bigg) &=&	\frac{1}{r_c^3 \gamma} \frac{\partial^2 \delta u(\gamma,t)}{\partial \gamma^2}Y_0^0(\theta,\phi)\nonumber\\
&=&\frac{Y_0^0}{r_c}\bigg[\Big(\frac{1}{r_c^2 \gamma}\frac{\partial^2}{\partial \gamma^2}\gamma\Big)\frac{\delta u(\gamma,t)}{\gamma}\bigg]\;\;\;\;\;\;\;\;\nonumber\\
&=& \frac{Y_0^0}{r_c} \nabla^2 \bigg[\frac{\delta u(\gamma,t)}{\gamma}\bigg].
\end{eqnarray}
The angular dependence of \eref{eqn:AppRadialPerturbationEqUns} can then be eliminated; dividing by $Y_0^0$ and multiplying by $r_c $,
\begin{eqnarray}
\bigg(-i\hbar \frac{\partial}{\partial t}-\frac{\hbar^2}{2m_b}\nabla^2 + m_bV_0\bigg) &&\frac{\delta u(\gamma, t)}{\gamma}\nonumber\\ = -m_b \delta&&V\frac{u_{00}(\gamma)}{\gamma} e^{-i E_{00}t/\hbar},	\;\;\;\;\;\;
\label{eqn:appRadialPerturbationEq}
\end{eqnarray}
which is coupled to the induced density perturbation given by \eref{eqn:appDensityWF}:
\begin{eqnarray}\label{app:CoupledDensity}
\delta \rho = \bigg(\frac{M_\text{soliton}}{4\pi r_c^2}\bigg)\times 2\frac{u_{00}}{\gamma}\frac{|\delta u(\gamma,t)|}{\gamma}.
\end{eqnarray}

For convenience, we shift the zero-level of $V_0$ such that $\psi_0$ has zero frequency
\begin{eqnarray}	
E_{00,\text{shifted}} \eee E_{00}-E_{00} = 0
\end{eqnarray}	
and zero phase; thus, $\psi_{000}$ (and $u_{00}$ by construction) is a real function. Since $\delta V$ is a real function, the RHS of \eref{eqn:appRadialPerturbationEq} is real; solutions of the form $f(r) e^{-i (\omega t + \phi)}$ for $\delta u(\gamma, t)$ is inexorably excluded. We adopt the following ansatz
\begin{eqnarray}
\delta u(\gamma, t) \eee \cos(\omega_{p} t+ \phi)F(\gamma) + i \sin(\omega_{p} t+ \phi)G(\gamma),\;\;\;\;\;\;\;\;\;\;\label{eqn:VPParametrization}
\end{eqnarray}
where both $F(\gamma)$ and $G(\gamma)$ are real functions, and
\begin{eqnarray}
\omega_{p}\eee \frac{E_p}{\hbar},
\end{eqnarray}
denotes the angular frequency of density perturbation. In this fluid representation of quantum system, we expect $E_p > 0$ that corresponds to an oscillation timescale of $2\pi/\omega_p$, while $E_p<0$ indicates instability. The perturbed Schr\"{o}dinger equation becomes
\begin{eqnarray}
\begin{cases}\label{app:FGCondition}
\hbar\omega \frac{G(\gamma)}{\gamma} - \Big( \frac{\hbar^2}{2m_b} \nabla^2 - m_bV_0\Big)\frac{F(\gamma)}{\gamma}=-m_b\delta \widetilde{V} \frac{u_{00}}{\gamma},\;\;\;\;\;\;\\
\hbar\omega \frac{F(\gamma)}{\gamma}- \Big( \frac{\hbar^2}{2m_b} \nabla^2 - m_bV_0\Big)\frac{G(\gamma)}{\gamma}=0,
\end{cases}
\end{eqnarray}
where we have introduced
\begin{eqnarray}
\begin{cases}
\delta V = \cos(\omega_{p} t+ \phi)\delta \widetilde{V},\\
\delta \rho =\cos(\omega_{p} t + \psi) \delta \widetilde{\rho}.
\end{cases}
\end{eqnarray}
The dependence of $G(\gamma)$ can be further eliminated:
\begin{eqnarray}
\hbar^2\omega_{p}^2 \frac{F(\gamma)}{\gamma}=\bigg( \frac{\hbar^2}{2m_b} \nabla^2 - m_bV_0\bigg)^2\frac{F(\gamma)}{\gamma}&&\nonumber\\-\bigg( \frac{\hbar^2}{2m_b} \nabla^2 - m_bV_0\bigg)m_b&&\delta \widetilde{V}\frac{u_{00}}{\gamma} \label{eqn:Ffunction},\;\;\;\;\;\;\;
\end{eqnarray}
where
\begin{eqnarray}
-\bigg( \frac{\hbar^2}{2m_b} \nabla^2 - m_bV_0\bigg) m_b\delta \widetilde{V} \frac{u_{00}}{\gamma}=-\bigg(\frac{\hbar^2}{2m_b}\nabla^2 m_b\delta \widetilde{V}\bigg)\frac{u_{00}}{\gamma}\nonumber\\- \bigg[\Big( \frac{\hbar^2}{2m_b} \nabla^2 - m_bV_0\Big)\frac{u_{00}}{\gamma}\bigg]m_b \delta \widetilde{V} - \frac{\hbar^2}{m_b} \nabla m_b\delta\widetilde{V}\times \nabla \frac{u_{00}}{\gamma}.\nonumber \label{eqn:SecondTerm}\\\;
\end{eqnarray}
The middle term on the RHS of \eref{eqn:SecondTerm} vanishes, as the value of gravitational potential at the reference point has been shifted such that $E_{00,\text{shifted}} = 0$.

Given the Laplacian and the gradient of perturbed potential
\begin{eqnarray}
\begin{cases}
\nabla^2 \delta \widetilde{V} = \frac{4\pi G \delta \rho}{\cos(\omega_{p} t +\phi)} = 4\pi G \delta \widetilde{\rho},\\
\nabla \delta \widetilde{V}\Big|_\gamma= \Big(4\pi G\frac{r_c}{\gamma^2}\int_0^\gamma d\gamma' \gamma'^2\delta \widetilde{\rho} \Big)\hat{\mathbf{r}},
\end{cases}
\end{eqnarray}
the first term on the RHS of \eref{eqn:SecondTerm} reads
\begin{eqnarray}
-\bigg(\frac{\hbar^2}{2m_b}\nabla^2 m_b\delta \widetilde{V}\bigg)\frac{u_{00}}{\gamma} - \frac{\hbar^2}{m_b} \nabla m_b\delta\widetilde{V} \times \nabla&& \frac{u_{00}}{\gamma}\nonumber\\
=-2\pi G\hbar^2\times\delta \widetilde{\rho} \frac{u_{00}}{\gamma}- \frac{\hbar^2}{m_b} \nabla m_b\delta\widetilde{V}\times&& \nabla \frac{u_{00}}{\gamma},\;\;\;\;\;\;\;\;
\end{eqnarray}
where first term on the RHS can be written as
\begin{eqnarray}	
-\bigg(\frac{\hbar^2}{2m_b}&&\nabla^2 m_b\delta \widetilde{V}\bigg)\frac{u_{00}}{\gamma}=-2\pi G\hbar^2\times\delta \widetilde{\rho} \frac{u_{00}}{\gamma}\nonumber\\=&&
-\bigg(\frac{G\hbar^2}{2 r_c^2}\bigg)M_\text{soliton}\times 2 \bigg(\frac{u_{00}}{\gamma}\bigg)^2 \bigg(\frac{|\delta u(\gamma,t)|}{\gamma\cos(\omega_p t+ \phi)}\bigg)\nonumber\\
=&& -11.6 G\hbar^2\rho_c r_c\times \bigg(\frac{u_{00}}{\gamma}\bigg)^2\frac{F(\gamma)}{\gamma},\;\;\;\;
\end{eqnarray}
and similarly second term on the RHS can be expanded as
\begin{eqnarray}	
 -&& \frac{\hbar^2}{m_b}\nabla m_b\delta\widetilde{V}\times \nabla \frac{u_{00}}{\gamma},\\
 =&&-4\pi G\nonumber\hbar^2r_c\Big(\frac{1}{\gamma^2}\int_0^\gamma d\gamma' \gamma'^2\delta \widetilde{\rho}\hat{\mathbf{r}}\Big) \nabla \frac{u_{00}}{\gamma} \\
=&& - 23.2 G\hbar^2\rho_c r_c^2\Big(\frac{1}{\gamma^2}\int_0^\gamma d\gamma'u_{00}F(\gamma')\hat{\mathbf{r}}\Big) \nabla \frac{u_{00}}{\gamma}.\nonumber
\end{eqnarray}	
The first-order perturbation equation (\eref{eqn:Ffunction}) then reduces to
\begin{eqnarray}\label{eqn:PerturbationEq}
&&\omega_p^2 \frac{F(\gamma)}{\gamma}= \Big(\frac{\hbar}{2m_b} \nabla^2 - \frac{m_b}{\hbar}V_0\Big)^2 \frac{F(\gamma)}{\gamma}-11.6G\rho_c r_c\\&&\times \Bigg\{
\bigg(\frac{u_{00}}{\gamma}\bigg)^2 \frac{F(\gamma)}{\gamma} + 2r_c\bigg[\frac{1}{\gamma^2}\int_0^\gamma d\gamma'u_{00}F(\gamma')\hat{\mathbf{r}}\bigg]\nabla \frac{u_{00}}{\gamma}\Bigg\}.\;\;\;\;\;\:\:\nonumber
\end{eqnarray}

\subsection{Variational Method}\label{app:VariationalMethod}

Consider a normalized combination of radial wavefunctions of energy levels $(n, 0), \forall n \in \mathbb{Z}^{0+}$ as the trial function
\begin{eqnarray}
F(\gamma) = c_{10} u_{10} + c_{20} u_{20} + ... + c_{n0} u_{n0},
\end{eqnarray}
where $\sum_{i=1}^{n} |c_{i0}|^2 = 1$, and $E_{i0,\text{shifted}} \eee E_{i0} - E_{00}$. The orthogonality and normalization condition give
\begin{eqnarray}
\int_0^\infty d\gamma F(\gamma) \bigg[\Big(\frac{\hbar}{2m_b}\nabla^2 - &&\frac{m_b}{\hbar}V_0\Big)^2F(\gamma)\bigg]\nonumber \\\;\;\;\;\;\;\;\;=r_c^{-1}&&\Big(\frac{\sum_{i=1}^{n} |c_i|^2 E_{i0,\text{shifted}}}{\hbar}\Big)^2.\;\;\;\;\;\;\;
\end{eqnarray}
The (radial) ground-state wavefunction $u_{00}$ is excluded in the analysis, as such a state induces no perturbation in the static potential. To normalize energy to $m_b V(0;\rho_c, m_b)$ (\eref{eqn:SolitonPotentialCenter}), we introduce the following dimensionless coefficient
\begin{eqnarray}
\frac{11.6 G \rho_c \hbar^2}{\big[m_b V_0(0;\rho_c, m_b)\big]^2} &=&0.536.
\end{eqnarray}
The timescale of density perturbations can then be readily estimated by multiplying both sides of \eref{eqn:PerturbationEq} with $F(\gamma)/\gamma$ and integrating over $\gamma = \left[0,\infty\right)$, yielding
\begin{eqnarray}
\omega_{p}^2 = &&\frac{\sum_{i=1}^{n} |c_{i0}|^2 E_{i0,\text{shifted}}^2}{\hbar^2}\nonumber\\&&\;\;\;\;\;\;\;\;\;\;\;\;\;\;\;\;\;\;\;+\bigg[\frac{m_b V_0(0;\rho_c, m_b)}{\hbar}\bigg]^2 \mathcal{F}(F(\gamma)).\;\;\;\;\;\;
\end{eqnarray}	
Equivalently, we have
\begin{eqnarray}
\bigg(\frac{\hbar \omega_{p}}{m_b V_0(0; \rho_c, m_b)}\bigg)^2 = E_\text{eff}^2+\mathcal{F}(F(\gamma)),
\end{eqnarray}
where we denote the effective scaled energy as
\begin{eqnarray}
E_\text{eff}\eee \sqrt{\sum_{i=1}^{n} |c_{i0}|^2\bigg[\frac{ E_{i0,\text{shifted}}}{m_b V_0(0; \rho_c, m_b)}\bigg]^2},
\end{eqnarray}
and the strictly negative self-gravity contribution as
\begin{eqnarray}
\mathcal{F}(F(\gamma))	\eee -0.&&536 r_c^2\int_0^\infty d\gamma F(\gamma)\Bigg\{
\Big(\frac{u_{00}}{\gamma}\Big)^2 F(\gamma) \nonumber\\\;\:\:&&+ 2r_c \bigg[ \frac{1}{\gamma}\int_0^\gamma d\gamma'u_{00}F(\gamma')\hat{\mathbf{r}}\bigg]\nabla \frac{u_{00}}{\gamma}\Bigg\}.\;\;\;\;\;\;\;\;\;
\end{eqnarray}

\begin{figure}
	\includegraphics[width=\linewidth]{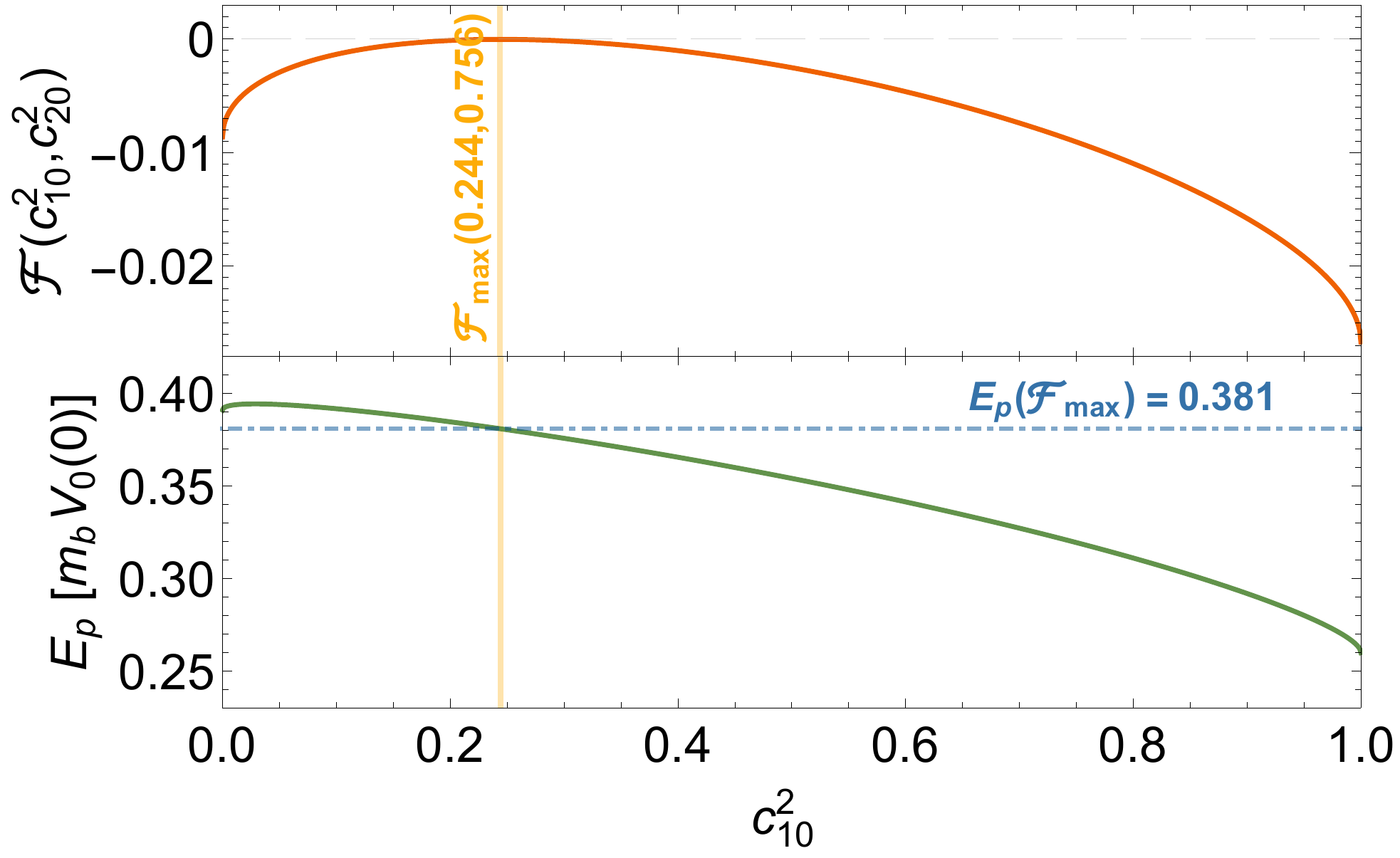}
	\caption{Values of $\mathcal{F}(F(\gamma))$ (top panel) and $E_p$ scaled to $m_b V(0; \rho_c, m_b)$ (bottom panel). We probe $0 \leq c_{10}^2 \leq 1$ and $c_{20}^2 = 1- c_{10}^2$, fixing $c_{30}= c_{40} =0$.}
	\label{fig:EpPlot}
\end{figure}

We apply the variational principle to estimate $E_p$ by extremizing $\mathcal{F}(F(\gamma))$ (maximizing in this case as $\mathcal{F}(F(\gamma)) < 0$)
\begin{eqnarray}
\frac{E_p}{m_b V_0(0; \rho_c, m_b)} = \sqrt{ E_\text{eff}^2+\mathcal{F}(F(\gamma))},
\end{eqnarray}
where all the shifted energy levels are positive ($E_{n0,\text{shifted}} = E_{n0} -E_{00} \geq 0, \forall n \in \mathbb{Z}^{+}$, where $E_{00} < 0$). We confine our choice of possible trial radial wavefunctions to $u_{10}, u_{20}, u_{30},$ and $u_{40},$ by noting that eigenstates $(n,0)$ become progressively less energetically favorable as $n$ increases. Without loss of generality, we work in a basis where $u_{n0} \in \mathbb{R}$ and choose $c_{n0} \in \mathbb{R}$ for $n \in \mathbb{Z}^{0+}$. A numerical scan across the $4$-dimensional $(c_{10}^2,c_{20}^2, c_{30}^2, c_{40}^2)$ parameter space locates a global maximum of $\mathcal{F}(F(\gamma))$ at $(c_{10}^2,c_{20}^2, c_{30}^2, c_{40}^2) = (0.244, 0.756, 0, 0)$ where
\begin{eqnarray}
\frac{E_p}{m_b V_0(0; \rho_c, m_b)} =0.381,
\end{eqnarray}
with the corresponding timescale
\begin{eqnarray}
\tau_p(\rho_c) &=&\frac{2\pi\hbar}{|E_p|} \simeq 53.1 \bigg(\frac{\rho_c}{\text{M}_\odot \text{pc}^{-3}}\bigg)^{-1/2} \text{ Myr}. \label{eqn:PerturbationPeriod}
\end{eqnarray}
Figure \ref{fig:EpPlot} shows the value of $\mathcal{F}(F(\gamma))$ in the top panel for $0 \leq c_{10}^2 \leq 1$ and $c_{20}^2 = 1- c_{10}^2$, fixing $c_{30}= c_{40} =0$. The maximum occurs at $(c_{10}^2, c_{20}^2) = (0.244, 0.756)$ (yellow). The bottom panel plots the value of $E_p(c_{10}^2,c_{20}^2)$.

Linear density perturbations (applicable for $\mathcal{C} \rightarrow 0$) offer an order-of-magnitude estimate of the genuine timescale of soliton density oscillations $\tau_\text{soliton}$, as $\delta \rho/\rho \sim \mathcal{O}(1)$. The nonlinear oscillation frequency tends to be lower than the linear frequency (e.g. nonlinear pendulum and our analysis in Appendix \ref{app:Resonance}). We compare the result with simulations and have $\tau_\text{soliton}(\rho_c)/\tau_p(\rho_c) \simeq 1.7$.

\section{Orbital Migration and Resonances}\label{app:HamiltonianPerturbation}
We are primarily interested in the particle mass range $m_b \lesssim 4.6 \times 10^{-22}$ eV; the SC is well embedded in the soliton ($r_\text{SC} \ll r_c$). The same analysis can in principle be extended to larger masses $m_b \gtrsim 10^{-21}$ eV, where the astrophysical constraint on $r_c$ would also depend on modeling of a background halo.

\subsection{Perturbation Hamiltonian}
The Hamiltonian formalism alternatively yields the same radial equation of motion. The Hamiltonian of a test star can be decomposed into a time-independent piece $H_0$ and time-dependent perturbation $H_1$ that accounts for the single-mode density fluctuations of a soliton within radius $r$:
\begin{eqnarray}\label{eqn:HamiltonianPert}
H_0 &=& \frac{m_\star \dot{r}^2}{2}+ \frac{l^2}{2 m_\star r^2}+
m_\star\bigg[\Big(-\frac{G M_\text{SC}}{r}\Big) + V_0(r; \rho_c, m_b)\bigg],\nonumber\\
H_1 &=& m_\star\Big[V_0(r;\rho_cs(t), m_b)-V_0(r;\rho_c, m_b)\Big],
\end{eqnarray}
where the conserved angular momentum reads $l = m_\star r^2 \omega_\text{SC} = m_\star r_\text{SC}^2\omega_{\text{SC},0}$. Any statistically symmetric density fluctuations of a soliton (or a halo) and subsequent change in potential exert no net force on a test star, thanks to the shell theorem. We therefore need not consider fluctuations outside of the sphere of radius $r$.

The $\rho_c$-varying soliton potential (\eref{eqn:VGeneric}) in $H_0+H_1$ reads
\begin{eqnarray}
V_0(r;\rho_cs(t), m_b) &&= -Gr_c^2\rho_c s(t)^{1/2} \times U\bigg(\frac{r}{r_cs(t)^{-1/4}}\bigg),\;\;\;\;\;\;\;\;\;
\end{eqnarray}
where on the RHS we have scaled out the time-dependence of both $r_c$ and $\rho_c$. The time evolution of orbital radius then follows
\begin{eqnarray}
\ddot{r} = \frac{\omega_{\text{SC},0}^2r_\text{SC}^4}{r^3}- G\bigg[\frac{M_\text{SC}}{r^2}-r_c\rho_c s(t)^{3/4} \times \frac{\partial U(\gamma)}{\partial \gamma}\bigg], \;\;\;\;\;\;\;\;\;\label{eqn:appTidalHeatingFullyGeneral}
\end{eqnarray}
which is equivalent to \eref{eqn:TidalHeatingMass}. To first order in $\gamma$, we have
\begin{eqnarray}
s(t)^{3/4}\frac{\partial U(\gamma)}{\partial \gamma} &=&s(t)^{3/4} \Big[-\frac{4\pi}{3}\gamma +\mathcal{O}(\gamma^2)\Big]\nonumber\\
&=&-\frac{4\pi}{3} s(t)^{3/4}\bigg(\frac{r}{r_c s(t)^{-1/4}}\bigg)+\mathcal{O}(\gamma^2)\nonumber\\
&=& -\frac{4\pi}{3}s(t) \bigg(\frac{r}{r_c }\bigg)+\mathcal{O}(\gamma^2).\;\;\;\;\;\;\;\;\;\;\;\;\;\;\;\;\;\;\;\;
\end{eqnarray}
This expression expectedly agrees with the equation of motion obtained by directly substituting the small-radius asymptotic expression of the soliton potential (\eref{eqn:PotentialSmallgApprox}) into \eref{eqn:HamiltonianPert}
\begin{eqnarray}\label{eqn:appTidalHeating2ndODE}
\ddot{r} = \frac{\omega_{\text{SC},0}^2r_\text{SC}^4}{r^3}- G\bigg\{\frac{M_\text{SC}}{r^2}+\frac{4\pi}{3} \rho_c r \Big[1+ \mathcal{C}\sin(\omega_\text{osc}t+\phi)\Big]\bigg\}.\nonumber\\
\end{eqnarray}

\section{Resonance Analysis}\label{app:Resonance}
\subsection{Linear Regime: Parametric Resonance}
\begin{figure}
	\includegraphics[width=\linewidth]{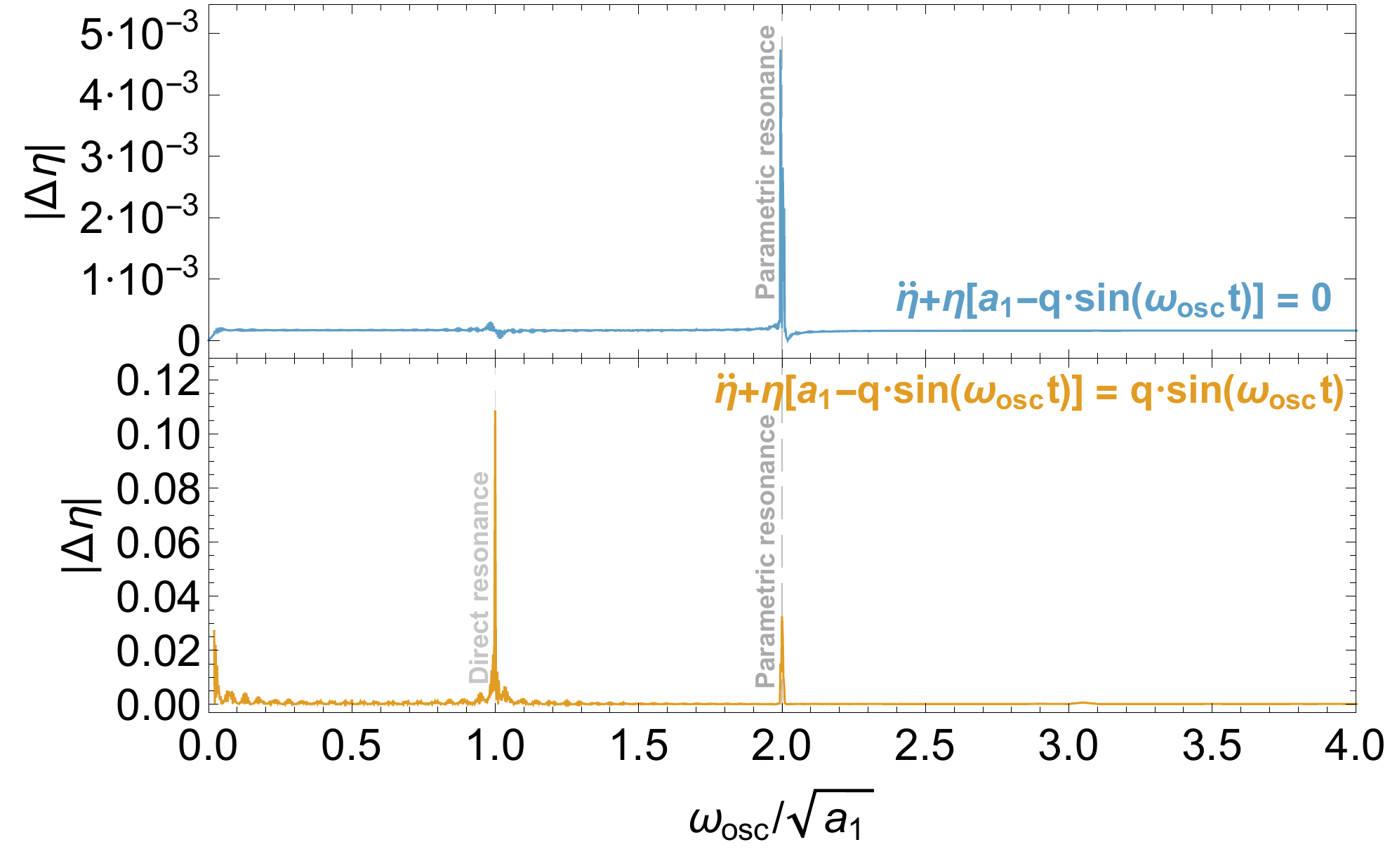}
	\caption{Value of $|\Delta \eta|$ by numerically solving \eref{eqn:ModifiedMathieuEqn} over 10 Gyr, fixing $m_b = 10^{-22}$ eV, $\mathcal{C} = 0.1$, and $\eta_0 = 10^{-2}$. \textit{Top:} The external driving term is excluded. The characteristic parametric resonance is observed at $\omega_{\text{osc}}/\sqrt{a_1} = 2$, as expected. \textit{Bottom:} The included driving force term gives rise to direct resonance at $\omega_{\text{osc}}/\sqrt{a_1} = 1$, while amplifying the magnitude of parametric resonance.}
	\label{fig:ParametricResonance}
\end{figure}

In the limit $\mathcal{C}\ll 1$, the time evolution of orbital radius can be written as $r \eee r_\text{SC} + r_1$, where $r_\text{SC}$ is the radius of an unperturbed orbit. Hence $\dot{r}_\text{SC} = \ddot{r}_\text{SC} = 0$, and
\begin{eqnarray}
\eta \eee \frac{r_1}{r_\text{SC}} \ll 1.\label{eqn:appetaDefinition}
\end{eqnarray}
\Seref{eqn:appTidalHeating2ndODE} to first order in $r_1$ reduces to a forced Mathieu equation (setting $\phi = 0$)
\begin{eqnarray}
\ddot{\eta} +\eta \bigg[a_1-q \sin(\omega_\text{osc}t)\bigg] = q \sin(\omega_\text{osc}t), \label{eqn:ModifiedMathieuEqn}
\end{eqnarray}
where
\begin{eqnarray}
\begin{cases}
a_1 \eee 3\omega_{\text{SC},0}^2 +G\Big(-2\frac{M_\text{SC}^{\le r_\text{SC}}}{r_\text{SC}^3}+\frac{4\pi}{3} \rho_c\Big),\\
q \eee \frac{4\pi}{3}G\mathcal{C}\rho_c.
\end{cases}
\end{eqnarray}
The RHS of \eref{eqn:ModifiedMathieuEqn} acts as an external driving source and gives rise to direct resonance at the frequency $\omega_\text{d.r.} = \sqrt{a_1}$.
Dropping this driving term, we recover the standard expression of Mathieu function, with parametric resonance at $\omega_\text{p.r.} = 2\sqrt{a_1}$.

Figure \ref{fig:ParametricResonance} plots the time-averaged change in $\eta$ as a function of $\omega_{\text{osc}}/\sqrt{a_1}$, by numerically integrating \eref{eqn:ModifiedMathieuEqn}. The top panel excludes the driving term; parametric resonance is identified at $\omega_\text{osc}/\sqrt{a_1} =2$. In the bottom panel, the forced Mathieu equation exhibits not only enhanced parametric resonance in magnitude, but direct resonance at $\omega_{\text{osc}}/\sqrt{a_1} = 1$.

\subsection{Nonlinear Perturbation}

\begin{figure}
	\includegraphics[width=\linewidth]{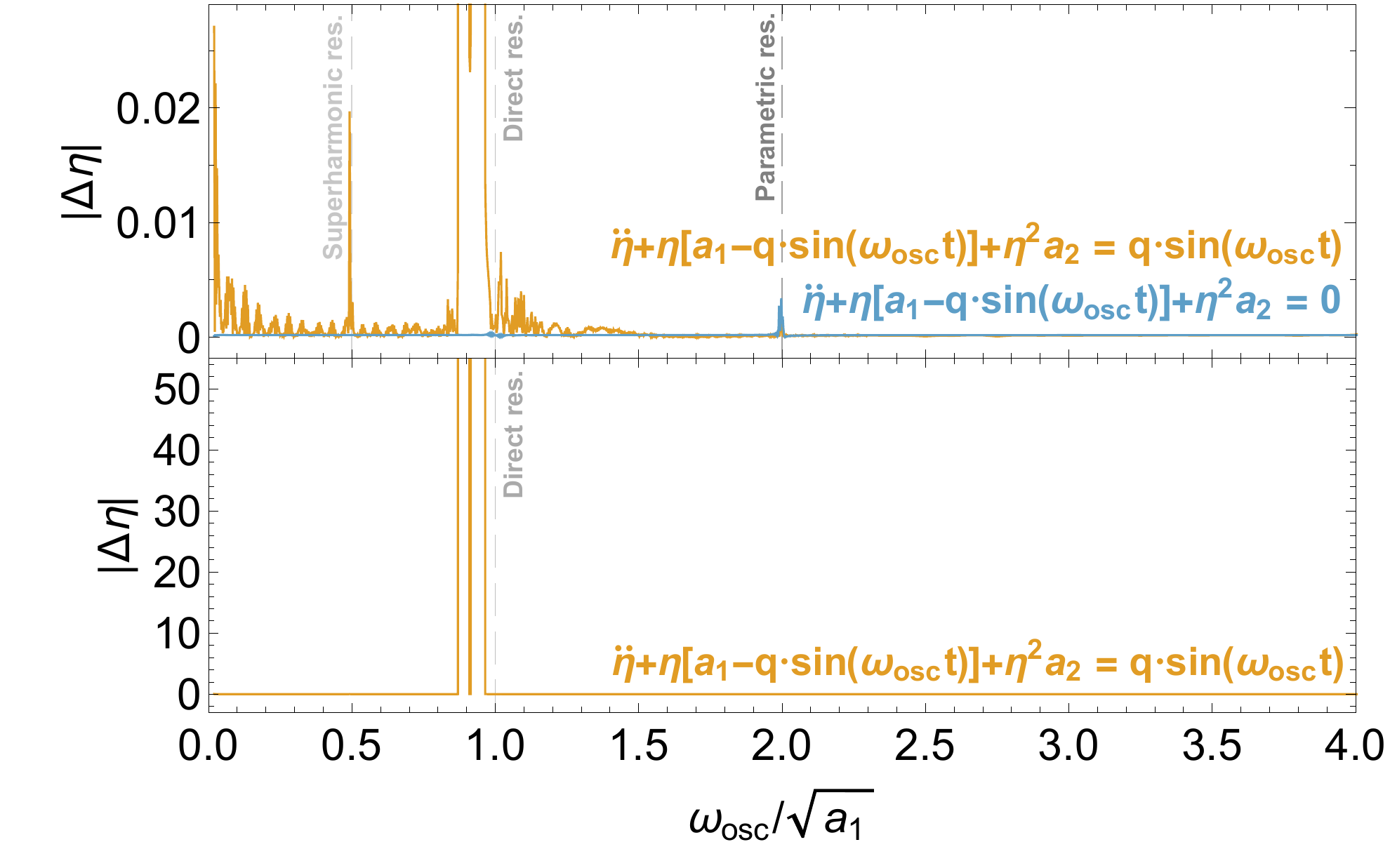}
	\caption{Value of $|\Delta \eta|$ by integrating \eref{eqn:SecondOrderCorrection} over 10 Gyr, fixing $m_b = 10^{-22}$ eV, $\mathcal{C} = 0.1$, and $\eta_0 = 10^{-2}$. The magnitude of direct resonance is greatly enhanced by the second-order correction, and the effect of parametric resonance becomes negligible in comparison. \textit{Top:} The external driving term is included/excluded (yellow/blue). \textit{Bottom:} The y-axis is rescaled to demonstrate $|\Delta \eta| \gg \mathcal{O}(1)$ for direct resonance.}
	\label{fig:ParametricResonanceNonLinear}
\end{figure}

The parametric behavior of \eref{eqn:ModifiedMathieuEqn} alters dramatically when higher-order corrections are introduced. Collecting terms in \eref{eqn:appTidalHeating2ndODE} up to the second-order in $\eta$, we have
\begin{eqnarray}
\ddot{\eta} +\eta \bigg[a_1-q \sin(\omega_\text{osc}t)\bigg]+\eta^2a_2+\mathcal{O}(\eta^3)&=&q \sin(\omega_\text{osc}t), \label{eqn:SecondOrderCorrection}\nonumber\\\;
\end{eqnarray}
where we have introduced
\begin{eqnarray}
a_2 \eee -6\omega_{\text{SC},0}^2 +G\bigg(3\frac{M_\text{SC}^{\le r_\text{SC}}}{r_\text{SC}^3}\bigg).
\end{eqnarray}

The top panel of Fig. \ref{fig:ParametricResonanceNonLinear} shows the case with (without) external driving term in yellow (blue). Direct resonance increases noticeably in magnitude, albeit shifted slightly towards lower frequency ($\omega_\text{d.r.}/\sqrt{a}<1$); the prominence of parametric resonance is heavily suppressed by nonlinear responses. We rescale the y-axis in the bottom panel to demonstrate that direct resonance yields
\begin{eqnarray}
|\Delta{\eta}| = \bigg|\frac{r_1}{r_\text{SC}}\bigg| \gg \mathcal{O}(1).
\end{eqnarray}

To account for the nonlinear effects of higher-order corrections, we note that direct resonance takes place at $\omega_\text{osc}/\sqrt{a_\text{eff}}=1$, where $a_\text{eff}$ is a of function $a_1, a_2, ...$. Given $a_\text{eff} = a_1$ in the first-order expansion of $\eta$ and the standard assumption that higher-order contributions $a_i$ in $a_\text{eff}$ for $i \in \mathbb{Z}^{+}$ become progressively negligible as $i$ increases, the ansatz leads to the following linearization
\begin{eqnarray}
a_\text{eff} \simeq a_1 +0.103 a_2 +\mathcal{O}(\eta^3),
\end{eqnarray}
by matching $\omega_\text{d.r.}$ with the direct resonance band for a fiducial choice of $\mathcal{C} = 0.3$, and therefore
\begin{eqnarray}
\omega_\text{d.r.} &\simeq&\sqrt{a_1+0.103 a_2}. \label{eqn:DirectResonanceaeff}
\end{eqnarray}

The forced Mathieu equation by expanding to arbitrarily high order in $\eta$ reads
\begin{eqnarray}
\ddot{\eta} +\omega_\text{d.r.}^2\eta\Bigg[ \Big(\frac{a_1}{a_\text{eff}}-\frac{q}{a_\text{eff}}\sin(\omega_\text{osc}t)\Big)+\sum_{i=2}^{\infty}&&\Big(\frac{a_i}{a_\text{eff}}\Big) \eta^i\Bigg]\nonumber\\=q&& \sin(\omega_\text{osc}t),\;\;\;\;\;\;\;\;\; \label{eqn:appHigherOrderCorrection}
\end{eqnarray}
and the nonlinear responses thereof can be characterized by a series of secondary superharmonic resonances at frequencies
\begin{eqnarray}
\omega_{\text{s.r.},n} \simeq \frac{\omega_\text{d.r.}}{n} = \frac{\sqrt{a_\text{eff}}}{n}, \label{eqn:Superharmonics}
\end{eqnarray}
for $n \geq 2$ $(n \in \mathbb{Z}^{+})$ \cite{Muthusamy:2003, RamakrishnanSHR}.

\bibliography{MyBibTeX1}
\bibliographystyle{JHEP}
\end{document}